\documentclass[11pt,a4paper]{article}
\usepackage{latexsym,makeidx,syntonly,amsfonts,amssymb}
\newcommand{\sect}[1]{\setcounter{equation}{0}\section{#1}}
\newcommand{\subsect}[1]{\subsection{#1}}

\newcommand{\vs}[1]{\rule[- #1 mm]{0mm}{#1 mm}}
\newcommand{\hs}[1]{\hspace{#1 mm}}
\newcommand{\lbl}[1]{\label{eq:#1}}
\newcommand{\rf}[1]{(\ref{eq:#1})}
\newcommand{\nn}{\nonumber}
\newcommand{\Ch}{{\cC^{(2)}}}

\newcommand{\be}{\vs{2}\begin{equation}}
\newcommand{\ee}{\vs{2}\end{equation}}

\newcommand{\bea}{\begin{eqnarray}}
\newcommand{\ena}{\end{eqnarray}}
\newcommand{\nnbea}{\begin{eqnarray*}}
\newcommand{\nnena}{\end{eqnarray*}}

\newcommand{\lra}{\ \longrightarrow\ }

\newcommand{\ovl}[1]{\overline{#1}}

\newcommand{\dps}{\displaystyle}

\newcommand{\bz}{{\ovl{z}}}

\newcommand{\zbz}{(z,\bz)}

\newcommand{\z}{(z)}
\newcommand{\zer}[1]{\stackrel{\circ}{#1}}

\newcommand{\wbw}{(w,\bar{w})}

\newcommand{\cW}{{\cal W }}

\newcommand{\sS}{{\cal S }}
\newcommand{\cK}{{\cal K }}
\newcommand{\cF}{{\cal F }}
\newcommand{\cC}{{\cal C }}

\newcommand{\cA}{{\cal A }}
\newcommand{\cD}{{\cal D }}

\newcommand{\cT}{{\cal T }}

\newcommand{\cM}{{\cal M }}

\newcommand{\cV}{{\cal V }}
\newcommand{\cB}{{\cal B }}
\newcommand{\cX}{{\cal X }}

\newcommand{\cZ}{{\cal Z }}

\newcommand{\cJ}{{\cal J }}

\newcommand{\prt}{\partial}
\newcommand{\prtz}{\partial_z}
\newcommand{\prtbz}{\partial_{\ovl{z}}}

\newcommand{\bprt}{\ovl{\partial}}

\newcommand{\mub}{\ovl{\mu}}

\newcommand{\ZBZ}{(Z,\ovl{Z})}

\newcommand{\zbzp}{(z',\ovl{z}')}

\newcommand{\zp}{{z}'}
\newcommand{\bzp}{\ovl{z}'}

\newtheorem{statement}{Statement}[section]

\begin{document}


\pagestyle{empty}

\font\fifteen=cmbx10 at 15pt
\font\twelve=cmbx10 at 12pt

\begin{titlepage}

\begin{center}
\renewcommand{\thefootnote}{\fnsymbol{footnote}}

{\twelve Centre de Physique Th\'eorique\footnote{
Unit\'e Propre de Recherche 7061, et FRUMAM/F\'ed\'eration de Recherches 2291
}, CNRS Luminy, Case 907}

{\twelve F-13288 Marseille -- Cedex 9}

\vskip 2cm

{\fifteen Induced quantum gravity on a Riemann Surface}

\vskip 1.5cm

\setcounter{footnote}{0}
\renewcommand{\thefootnote}{\arabic{footnote}}

{\bf G. BANDELLONI} $^a$\footnote{e-mail : {\tt beppe@genova.infn.it}}
\hskip .2mm  and \hskip .7mm {\bf S. LAZZARINI} $^b$\footnote{and also
Universit\'e de la M\'editerran\'ee, Aix-Marseille II. e-mail : {\tt
sel@cpt.univ-mrs.fr} }\\[6mm]
$^a$ \textit{Dipartimento di Fisica
dell'Universit\`a di Genova,}\\
{\it Via Dodecaneso 33, I-16146 GENOVA, Italy}\\
 and \\
{\it Istituto Nazionale di Fisica Nucleare, INFN, Sezione di Genova}\\
{\it via Dodecaneso 33, I-16146 GENOVA, Italy}\\[4mm]
$^b$ {\it Centre de Physique Th\'eorique, CNRS Luminy, Case 907,}\\
{\it F-13288 MARSEILLE Cedex, France}
\end{center}

\vskip 1.cm

\centerline{\bf Abstract}
Induced quantum gravity dynamics built over a Riemann
surface is studied in arbitrary dimension.
Local coordinates on the target space are given by means of the
Laguerre-Forsyth construction.
A simple model is proposed and pertubatively quantized. In
doing so,
the classical $\cW$-symmetry turns out to be preserved on-shell at any
order of the 
$\hbar$ perturbative expansion. As a main result, due to quantum
corrections, the target coordinates acquire a non-trivial character.

\vskip 1.5cm

\noindent 1998 PACS Classification: 11.10.Gh - 11.25 hf - 03.70


\indent

\noindent{CPT--2002/P.4415}, \hfill{Internet : {\tt www.cpt.univ-mrs.fr}}

\end{titlepage}

\renewcommand{\thefootnote}{\arabic{footnote}}
\setcounter{footnote}{0}
\pagestyle{plain}
\setcounter{page}{1}

\newpage

\sect{Introduction}

The relationship between quantum world and gravitational force is
nowadays one of the important issues in theoretical physics, and from
several decades, it has been widely debated in the community.
Indeed, in the classical theory, the dynamics of gravitation is
strictly linked with space-time attributes only, so that any
quantum extension should investigate  the interactions between
matter fields and space-time which ought to arise at very small
length scales \cite{Wheeler:1957mu,Hawking:1978zw,Crane:1986yg}.

 A local  field theory  describes the dynamics of physical phenomena
with the help of a local coordinate system
\cite{Descartes1600,Descartes16001}; as is well-known a physical
theory turns out to be valid if  any choice of the coordinate system
leaves unaffected the physics.  For instance string and (Mem)brane
theories greatly take advantage of these aspects.  But, even if these
facts are widely exploited from the mathematical point of view, some
new requirements (as non-commutativity) raise new questions on the
role of the background with respect to the physics investigations.

In particular the physical locality
requirements must be re-examined at the light of the need  that any theory
must also be globally
defined. We here re-propose the well-known
Laguerre-Forsyth
construction for the coordinates \cite{A.R.Forsyth:Forsyth}, already used
\cite{DiFrancesco:1991qr,Bauer:1991ai,Gieres:1993sd,Govindarajan:1995wm,Zucchini:1993xm,Govindarajan:1995wm,BaLa:2001wa}
in  bidimensional conformal field models.
This scheme, at the same time, exploits the benefits of
two-dimensional complex space, and describes manifolds of arbitrary
dimensions, by preserving some prescriptions required
both by Mathematics and Physics, that is, each quantity must be well
defined from the geometrical point of view. 

Then we avail of this construction to build a simple physical
model in a Lagrangian framework, and to extend it at the Quantum
level.

\indent

The Laguerre-Forsyth construction stems from the remark that the
$s$ linearly independent solutions of the $s$-th order globally
defined linear differential equations on a Riemann surface
\cite{G.Bol:bol} endowed with an complex analytic atlas with local
corrdinates $\zbz$, 

\begin{eqnarray}
L_s f_i\zbz\equiv \sum_{j=0}^s a_{(j)}^{(s)}\zbz\,
\prt^{(s-j)}f_i\zbz=0, \qquad i=1,...,s
\lbl{equazlf}
\end{eqnarray}
with the normalization choice
\begin{eqnarray}
a_{(0)}^{(s)}\zbz=1,
\lbl{normalization}
\end{eqnarray}
make sense only if they are scalar densities of conformal weight
$\frac{1-s}{2}$. The space of such densities will be denoted by
$\cV_{\frac{1-s}{2}}$. So it is possible to construct $(s-1)$
independent scalar objects (for example $\frac{f_i\zbz}{f_1\zbz})$
which define an $(s-1)$ dimensional  space.
More generally the operator $L_s$ maps scalar densities of conformal
weight $\frac{1-s}{2}$ into those of weight $\frac{1+s}{2} $.
Accordingly if
 \begin{eqnarray}
 g\zbz\in \cV_{\frac{1+s}{2}}, \quad \mbox{and} \quad
 f\zbz,f_i\zbz\in \cV_{\frac{1-s}{2}}
\lbl{imageii}
 \end{eqnarray}
 the image of the differential operator $L_s$ can be constructed by
means of a given basis of solutions of \rf{equazlf} by writing
\begin{eqnarray}
g\zbz=L_s\zbz f\zbz = \left|\begin{array}{cccc}
f\zbz&f_1\zbz&\cdots&f_s\zbz\\
\prt f\zbz&\prt f_1\zbz&\cdots&\prt f_s\zbz\\
\vdots&\vdots &\ddots& \vdots \\
\prt^{(s-1)}f\zbz&\prt^{(s-1)}f_1\zbz&\cdots&\prt^{(s-1)}f_s\zbz \\
\prt^{(s)} f\zbz&\prt^{(s)}f_1\zbz&\cdots&\prt^{(s)}f_s\zbz \\
\end{array}
\right|
\lbl{system}
\end{eqnarray}
where the coefficients $a^{(s)}_{(j)}\zbz$ can be evaluated from the
minors of the previous matrix Eq\rf{system}~\cite{DiFrancesco:1991qr}
\begin{eqnarray}
a^{(s)}_j\zbz =
\left|\begin{array}{cccc}
\prt^{(s)} f\zbz&\prt^{(s)}f_1\zbz&\cdots&\prt^{(s)}f_s\zbz \\
\vdots&\vdots &\ddots& \vdots \\
\prt^{(j+1)}f\zbz&\prt^{(j+1)}f_1\zbz&\cdots&\prt^{(j+1)}f_s\zbz \\
\prt^{(j-1)}f\zbz&\prt^{(j-1)}f_1\zbz&\cdots&\prt^{(j-1)}f_s\zbz \\
\vdots&\vdots &\ddots& \vdots \\
\prt f\zbz&\prt f_1\zbz&\cdots&\prt f_s\zbz\\
f\zbz&f_1\zbz&\cdots&f_s\zbz
\end{array}
\right|.
 \lbl{coefficients}
\end{eqnarray}

Introducing now a $s\times s$-matrix ${\bf F}\zbz$ defined as
\begin{eqnarray}
{\bf F}\zbz\equiv\left(\begin{array}{ccc}
f_1\zbz&\cdots&f_s\zbz\\
\prt f_1\zbz&\cdots&\prt f_s\zbz\\
\vdots &\ddots& \vdots \\
\prt^{(s-1)}f_1\zbz&\cdots&\prt^{(s-1)}f_s\zbz \\
\end{array}
\right)  \lbl{boldf}
\end{eqnarray}
the normalization condition\rf{normalization} is simply written as
\begin{eqnarray}
\det{\bf F}=1 \lbl{normalization1}
\end{eqnarray}
Now if the $s$ linearly independent solutions are constructed as
\begin{eqnarray}
f_j\zbz =  Z^{(j-1)}\zbz\,   \omega^{-\frac{1}{s}}\zbz,
 \quad\mbox{for}\quad j =1,..., s
\lbl{solutionsf}
\end{eqnarray}
with $Z^{(0)}\zbz=1$, and
$\omega\zbz$  the Wronskian
\begin{eqnarray}
\omega\zbz =\left|\begin{array}{ccc}
\prt Z^{(1)}\zbz&\cdots&\prt Z^{(s-1)}\zbz\\
 \vdots &\ddots& \vdots \\
\prt^{(s-1)} Z^{(1)}\zbz&\cdots&\prt^{(s-1)} Z^{(s-1)}\zbz
\end{array}
\right| \lbl{omega}
\end{eqnarray}
in order to preserve the normalization \rf{normalization}. Note
that the so introduced $s-1$ independent (Laguerre-Forsyth)
coordinates $Z^{(r)}\zbz$  specify, for each point of the
two-dimensional $\zbz$ space, a $(s-1)$-uple of local complex
coordinates in some complex target manifold of $s-1$ complex
dimensions. This
 definition assigns to the coordinates a scalar field label; the role of
 the two-dimensional $\zbz$ background must be exploited both from
 the mathematical and (mostly) the
physical point of view.
Due to \rf{coefficients} the
coefficients $a^{(s)}_j\zbz$ can be written
\cite{Gunning:1996up,R.C.Gunning:1967} as well-defined
functions of the $ Z^{(r)}\zbz$ functions and their $z$-derivatives
and as shown before the solutions $f_i\zbz$ as well.

Consider now, for a fixed point on the Riemann surface 
with local complex coordinates $\zbz$, a change of 
local complex coordinates in the target manifold
\begin{eqnarray}
\biggl(Z^{(1)}\zbz,\cdots, Z^{(s-1)}\zbz\biggr)\lra
\biggl(Z'^{(1)}\zbz,\cdots, Z'^{(s-1)}\zbz\biggr)
\end{eqnarray}
with the primed analogues of Eq\rf{solutionsf}. This gives rise to
new solutions $f'_i\zbz$ of an analogous differential equation
\rf{equazlf} in which the coefficients have been transformed but
$a_0=1$ and $a_1=0$ are kept fixed. Accordingly this allows one to construct a
new matrix ${\bf F'}\zbz$  whose structure repeats \rf{boldf} and
\rf{normalization1}.

Therefore the change of local complex coordinates on the target
 manifold induces a SL$(s)$ coordinate mapping:
\begin{eqnarray}
{\bf F'}\zbz &=&{\bf {M}}\zbz {\bf F}\zbz
\lbl{transfprime}
\end{eqnarray}
 with
\begin{eqnarray}
\det{\bf {M}}\zbz=1 \lbl{detMeq1}.
\end{eqnarray}
In terms of the coordinates $Z^{(j)}$ the mapping Eq \rf{transfprime}
explicitly writes
\begin{eqnarray}
&&Z'^{(j')}\zbz = Z^{(j)}\zbz \nn\\
&& +\ \frac{{\dps \sum_{r=1}^s
{\bf M}_{1,r}\zbz \biggl( \sum_{m=1}^s \left( \begin{array}{c}
r-1\\ m
\end{array}\right) \prt^m
Z^{(j)}\zbz\prt^{r-1-m}\omega^{-1/s}\zbz\biggr) }} {{\dps
\sum_{\ell=1}^s{\bf M}_{1,\ell}\zbz
\prt^{\ell-1}{\omega}^{-1/s}\zbz}}.
\lbl{Zfintrans}
\end{eqnarray}
This mapping is performed, as said before, for a fixed (but
arbitrary) point with local complex coordinates $\zbz$; for this
reason we must extend the former to the whole Riemann surface endowed
with a background 
complex structure defined by the system of local complex analytic
coordinates $\zbz$. 

To sum up these conclusions we formulate the following:
\begin{statement}
The Laguerre-Forsyth construction induces a mapping from each point
of the two-dimensional $\zbz$ plane to the $(s-1)$-dimensional
target space $ \biggl(Z^{(1)}\zbz,\cdots, Z^{(s-1)}\zbz\biggr) $.
The physical requirement for the equivalence of all the patching in
the $(s-1)$-dimensional target space
 requires the double exigency that they must be
independent on the choice of the point of the $\zbz$ plane from
which we start the construction, and each  reparametrization  as
in Eq \rf{Zfintrans} will describe the symmetry principle of each
dynamical model embedded in these spaces. As will be seen in the
following, its infinitesimal counterpart generates a $\cW_{s}$-symmetry.
\lbl{covariance}
\end{statement}

 Going on, let us introduce now the Maurer-Cartan form whose matrix valued
 components are
\begin{eqnarray}
\cA_z\zbz&=& {\prt {\bf  F}\zbz}{({\bf  F}\zbz)}^{-1}
\lbl{Az}\\
\cA_{\bz}\zbz&=&{\bprt {\bf  F}\zbz}{({\bf  F}\zbz)}^{-1}
\lbl{Abz}
\end{eqnarray}
with values in the $s\ell(s)$ Lie algebra. It automatically satisfies
the flatness structure equation (the so-called zero curvature
condition, see for instance \cite{Garajeu:1995jn})
which reads in components
\begin{eqnarray}
\prt \cA_{\bz}\zbz -\bprt \cA_z\zbz + \cA_\bz\zbz \cA_z\zbz
-\cA_z\zbz \cA_{\bz}\zbz=0 .\lbl{nullcurvature}
\end{eqnarray}
The construction of the matrix ${\bf{F}}\zbz$ by means of the
solutions of Eq\rf{equazlf}, gives a $(1,0)$-component,
$\cA_z\zbz$ which turns out to be written in the so-called
Drinfeld-Sokolov form. 

With the transformation Eq \rf{Zfintrans} the components in
Eq\rf{Az} Eq\rf{Abz} will transform by gauge transformations as:
\begin{eqnarray}
\cA_z'\zbz =(\prt{\bf {M}}\zbz){{{\bf {M}}\zbz})}^{-1} +{\bf
{M}}\zbz \cA_z\zbz {({\bf {M}}\zbz)}^{-1}\lbl{Azprime}\\
\cA_{\bz}'\zbz =(\bprt{\bf {\cM}}\zbz){{{\bf {M}}\zbz})}^{-1}
+{\bf {M}}\zbz \cA_\bz\zbz {({\bf
{M}}\zbz)}^{-1}\lbl{Abzprime}
\end{eqnarray}
We remark that for a general linear (rigid) reparametrization
\begin{eqnarray}
Z'^{(i)}\zbz = \sum_{j=1}^{s-1} M^{(i)}_{(j)}(\bz)
Z^{(j)}\zbz,\qquad |M(\bz)|\equiv \det M(\bz)\neq 0
\lbl{linear}
\end{eqnarray}
which is left out in Eq\rf{Zfintrans}, are relevant in our
treatment. It is easy to verify the density property of the wronskian
\begin{eqnarray}
\omega'\zbz=\omega\zbz |M(\bz)|.
\end{eqnarray}
Setting a new $s\times s$  matrix $M'(\bz)$ as:
\begin{eqnarray}
M'(\bz)= \left(\begin{array}{c} 1\\
 \hline\\[-4mm]
  0\end{array}\right|
\left.
\begin{array}{c}
0\\
\hline\\
[-4mm]
 M (\bz )  \end{array}
 \right)
\end{eqnarray}
we can define the homologous  $ f'_i\zbz $ functions  as:
\begin{eqnarray}
f'_i\zbz =  \frac{1}{{|M'(\bz)|}^{1/s}}
 \sum_j {M'(\bz)}_i^j f_j  \equiv \sum_{j=1}^s   {\cM(\bz)}_i^j f_j\zbz
\end{eqnarray}
where ${\cM(\bz)} $ is thus unimodular.

It is obvious that the $f'_i\zbz$ functions are still solutions of
equations \rf{equazlf} and it is easy to verify,
 by means of the construction \rf{coefficients} that the reparametrization
 in Eq\rf{linear} leaves the coefficients $a^{(s)}_j\zbz$ (and so the
matrix $\cA_z\zbz$)     invariant.
This shows in particular that a multiplicative renormalization of
the $Z^{(r)}\zbz$ fields never affect the matrix
elements $a^{(s)}_j\zbz$.

So with the help of two-dimensional conformal theory built on a
 Riemann surface, we can construct  coordinate transformations on a
 (target) complex space of arbitrary dimensions with well defined symmetry
 recipes. The main conceptual difference between our treatment and
 the mainstream one, is the privilege of the complex structure
 with respect to the metric properties of the space.

In this context we  shall construct a simple model, whose quantum
extension can be carried out in a perturbative way by a B.R.S quantization
approach.

The Laguerre-Forsyth construction will be the strategic trick of
our approach which enables to use the features of the
two-dimensional $\zbz$ space to build and investigate models in an
arbitrary number of dimensions.

The first constituent, we shall use as a probe, of our theoretical
laboratory is a field $\cX\zbz$ which must be scalar under any
coordinate transformation, and in particular under  the ones given by
Eq\rf{Zfintrans}.
Through this field it will be then possible to construct a well defined
Classical Lagrangian
invariant under the B.R.S. algebra induced by the $Z^{(r)}\zbz$
field transformations.
In particular, the $\cX\zbz$ equation of motion will define a
mass-shell  solution 
$\cX(Z^{(1)}\zbz,\cdots,
Z^{(s-1)}\zbz)$ which will describe the scalar field on a
$(s-1)$-dimensional target space.

We shall see that this Classical Lagrangian depends, as
regarding the $Z^{(r)}$ content, on their Beltrami coefficients only.

The Quantum extension of the model produces anomalies
 which spoil the symmetry. A counter-term machinery
 will compensate these anomalies, but will introduce new  pure gravitational
interactions at the quantum level. We shall show that  this
Lagrangian improvement will induce a kind of ``on mass-shell symmetry'' which
could be extended at any order of the perturbative expansion.

The quantum corrections to the model will generate an $\hbar$ order to the
$<Z^{(r)}\zbz Z^{(s)}\zbz>$ propagator whose 
ordering originates a non-commutative character of the
$Z^{(r)}\zbz$ fields-coordinates.
This idea is not new, and has already been proposed, within the $\cW$-algebra
framework, by many authors
\cite{Schoutens:1989tn,Schoutens:1990ja,Li:1990wx,Hull:1991ca,Schoutens:1991hx,Ceresole:1991hi,Schoutens:1991dq,Schoutens:1991sp,Hull:1991sa,Ooguri:1992by,Hull:1992hy,Schoutens:1992tm,Bouwknegt:1993wg,Hull:1993vj,Abud:1995,Ader:1996pr}.
But in these approaches the existence of the $\cW$-symmetry, does
not imply a space symmetry or any coordinate transformation
invariance in dimension greater than two.

In the approach we follow here (something similar can be found in
\cite{Bershadsky:1989mf}), the $\cW_s$-symmetry originates  from the
SL$(s)$-symmetry in eq \rf{transfprime} in a $(s-1)$-dimensional space
and not an induced symmetry coming form some OPEs. 
All the operators can be expressed by means of the $Z^{(r)}\zbz$
coordinates and their derivatives. Therefore the dynamical content 
considered in the paper has only to do with gravitational degrees of freedom
and not with some properties of the primary fields.

\indent

In Section 2 we build the algebra as infinitesimal transformations
of the $Z^{(r)}\zbz$ coordinates and we recall many of its properties.
 \vspace{0.5 cm}

In Section 3, using a scalar field as a probe, a
two-dimensional Lagrangian Field Theory model will be presented and
its quantum extension 
wil be performed only on-shell. The compensation of anomalies
produces a non trivial, and well-defined, pure gravitational
sector. When the scalar field equation of motion is solved the former
becomes a function of the $Z^{(r)}\zbz$ coordinates, and so, it turns
out to be
defined in an arbitrary dimensional space. This sector is
investigated by means of Ward identities, and the noncommutativity
of the $Z^{(r)}\zbz$ coordinates is put into evidence at the quantum level.

\indent

Some Appendices are devoted to some technicalities. In Appendix A we
calculate in a closed form the 
derivatives with order greater than $s$ of the functions $f_i\zbz$,
which become necessary to give a closed form of the B.R.S. algebra
of the ghost fields. In Appendix B, the compensation of anomalies
depending on the external fields
in the quantum extension of the model is carried out. Finally,
  in Appendix C the Ward
identities are specialized to the investigation of the $\cW_2$ and
$\cW_3$ cases, in order to exemplify our results.

\sect{$\cW$ Algebra in the $\cK$ ghosts}

Following \cite{Zucchini:1993xm}, the infinitesimal version of the
transformation \rf{transfprime} can be casted in a
B.R.S. formulation by introducing a ghost matrix $\cK\zbz$ :
\begin{eqnarray}
\delta_\cW{\bf  F}\zbz&=&\cK\zbz{\bf  F}\zbz,
\lbl{deltaF}
\end{eqnarray}
the nilpotency condition $\delta_\cW^2=0$ requires:
\begin{eqnarray}
\delta_\cW \cK\zbz =\cK\zbz \cK\zbz.
 \lbl{deltaw}
\end{eqnarray}
The ghost transformation contains many properties of the algebra at hand,
in particular:
\begin{eqnarray}
\delta_\cW \det{\cK\zbz}&=&Tr{\cK\zbz},     \det{\cK\zbz}
\lbl{deltadet}\\
Tr \cK^{2n}\zbz&=&0 \lbl{trk2n},\\
\delta_\cW Tr \cK^{2n+1}\zbz&=&0,\quad \lbl{deltak2np1} \forall
n=0,1\cdots
\end{eqnarray}
and from condition \rf{normalization1}
\begin{eqnarray}
\sum_{i=1}^s \cK_{i,i}\zbz=0. \lbl{trace}
\end{eqnarray}
So, from the very definition, we infinitesimally get as gauge transformations
\begin{eqnarray}
&&\delta_\cW  \cA_z\zbz =\prt \cK\zbz-\biggl[\cA_z\zbz ,\cK\zbz
\biggr], \lbl{deltaA}\\
&&\delta_\cW  \cA_\bz\zbz =\bprt \cK\zbz- \biggl[\cA_\bz\zbz
,\cK\zbz \biggr] . \lbl{deltaAb}
\end{eqnarray}
The BRS algebraic version of the coordinate transformations in
Eq \rf{Zfintrans} are derived from the first row of Eq \rf{deltaF}
and the definition \rf{solutionsf} as:
\begin{eqnarray}
\delta_\cW  Z^{(j)}\zbz &\equiv& \Lambda^{(j)}\zbz \nn\\
&=&
\sum_{\ell=1}^{s}\cK_{1,\ell}\zbz\sum_{h=0}^{\ell-1}
\left(\begin{array}{c}\ell-1\\
h\end{array}\right)\prt^{h} Z^{(j)}\zbz\,
 \frac{\prt^{\ell-1-h}{\omega^{-1/s}\zbz}}{\omega^{-1/s}\zbz}.
 \lbl{Zvariations0}
\end{eqnarray}
These coordinate transformations have not to be confused  with the
patching laws by holomorphic changes of charts   $z\lra w(z)$. The
latter, on the other hand, induce a gluing rule for a holomorphic
bundle of matrices with jet coordinates as entries, namely,
\begin{eqnarray}
{\bf F}\wbw =\Psi^{\frac{1-s}{2}}(z) {\bf F}\zbz \lbl{chartsF}
\end{eqnarray}
where the matrix valued holomorphic transition function is defined to be
\begin{eqnarray}
\Psi^{\frac{1-s}{2}}(z) = \Phi^{\frac{1-s}{2}}(z)^{-1}{}^{\mbox{\tt t}}
\lbl{psiprop}
\end{eqnarray}
where {\tt t} means transposition and where the invertible matrix
\begin{eqnarray}
\Phi^{\frac{1-s}{2}}_{\ell,k}(z) = \left\{ \begin{array}{lc} {\dps
\sum_{r=\ell}^k \frac{k!}{(k-r)!}
\prt_z^{k-r}\Big((w')^{\frac{1-s}{2}}\Big) \sum_{\footnotesize
\begin{array}{c}
a_1+\cdots+ n a_n= r\\
a_1+\cdots+  a_n= \ell \end{array} } \left(\prod_{n=1}^r
\frac{1}{a_n!} \left(\frac{w^{(n)}}{n!}\right)^{a_n}\right), } &
\ell \leq k\\ 0, & \ell > k
\end{array}\right.
\lbl{phi}
\end{eqnarray}
is in SL$(s,C)$ namely \cite{Gunning:1996up,R.C.Gunning:1967}
\begin{eqnarray}
\det \Phi^{\frac{1-s}{2}}(z) = \prod_{k=0}^{s-1}
\Phi^{\frac{1-s}{2}}_{k,k}(z) = 1. \lbl{detphi}
\end{eqnarray}
Under such a holomorphic change of charts our objects transform as:
\begin{eqnarray}
w'\, \cA_w \wbw &=& \biggl(
\prt_z\Psi^{\frac{1-s}{2}}(z) \biggr)\Psi^{\frac{1-s}{2}}(z)^{-1}
+{\Psi^{\frac{1-s}{2}}(z) }\cA_z\zbz\Psi^{\frac{1-s}{2}}(z)^{-1},
\lbl{changeconnect0}\\
\ovl{w}'
\cA_{\ovl{w}}\wbw &=& \Psi^{\frac{1-s}{2}}(z)
\cA_\bz\zbz \Psi^{\frac{1-s}{2}}(z)^{-1}, \lbl{changeconnect01}\\
\cK\wbw &=&
\Psi^{\frac{1-s}{2}}(z) \cK\zbz \Psi^{\frac{1-s}{2}}(z)^{-1}.
\lbl{changematrix0}
\end{eqnarray}
In particular note that:
\begin{eqnarray}
\cK_{1,\ell}\zbz &=& {(w')}^{\frac{1-s}{2}}\sum_{j\geq \ell}\cK_{1,j}\wbw 
\Psi_{\ell,j}^{\frac{1-s}{2}} (z), \quad
\ell=1,\cdots, s-1
\lbl{changeghosts0}\\
\cK_{1,s}\zbz &=& {(w')}^{{1-s}} \cK_{1,s}\wbw.
 \lbl{maximum}
\end{eqnarray}
So the $(1,s)$ entry of the ghost matrix $\cK\zbz$ transforms as a
tensorial density of conformal weight $s-1$.

Since the ${\bf F}_{(i,j)}\zbz$ matrix
 elements are derivatives of the $s$ solutions $f_i\zbz$,  the $
 \cK\zbz$ matrix elements
 are not all independent ; in fact for $j=0,\cdots, s-1$ :
\begin{eqnarray}
&&\delta_\cW \prt^j f_i\zbz \equiv\sum_{r=1}^s \cK_{j+1,r}\zbz
\prt^{r-1}f_i\zbz \nn\\ 
&&= \sum_{r=1}^{s}\prt^j\biggl[ \cK_{1,r}\zbz \prt^{r-1}f_i\zbz \biggr]\nn\\
&&= \sum_{r=1}^s \sum_{l=0}^j  \left(\begin{array}{c}
j\\l\end{array}\right)    \prt^{j-l}\cK_{1,1+(s-r)}\zbz \prt^{s+(l-r)}
f_i\zbz\nn\\ 
&&= \sum_{r=1}^s \sum_{l=0}^j  \left(\begin{array}{c} j\\l\end{array}
\right)    \prt^{j-l}\cK_{1,1+(s-r)}\zbz\sum_{k=1}^{s-1}\cF^{(r-l)}_{(k)}(s;\zbz) \prt^{s-k} f_i\zbz
\lbl{kdependence}
 \end{eqnarray}
 (the meaning of the $\cF^{(r-l)}_{(k)}(s;\zbz) $ fields, comes from
the fact that the derivatives of the $f_i\zbz$ functions of order
greater than $s-1$ can be expressed in terms of the ones of lower
orders. This aspect is explained in Appendix A).

For this reason we have  $ s-1$ independent $\Phi$-$\Pi$ charged
ghosts $\cK_{1,r}\zbz, r= 2,\cdots, s$.
We can state:
\begin{statement}
The $\cK_{j+1,m}\zbz;j>0$ are related to the $\cK_{1,m}\zbz$
by:
\begin{eqnarray}
&&\cK_{j+1,m}\zbz=   \Biggl[\sum_{r=1}^s \sum_{l=0}^j
\left(\begin{array}{c} j\\l\end{array} \right)
\prt^{j-l}\cK_{1,1+(s-r)}\zbz
\cF^{(r-l)}_{(s-m+1)}(s;\zbz)\Biggr].
\lbl{kstatement}
\end{eqnarray}
This formula holds also for $j=0$.
\end{statement}
So Eq\rf{trace} requires:
\begin{eqnarray}
&&\cK_{1,1}\zbz =-\sum_{j=1}^{s-1} \Biggl[\sum_{r=1}^s
\sum_{l=0}^j  \left(\begin{array}{c} j\\l\end{array} \right)
\prt^{j-l}\cK_{1,1+(s-r)}\zbz\cF^{(r-l)}_{(s-j)}(s;\zbz)\Biggr].
\lbl{k11}
\end{eqnarray}
The same holds for the two matrices $\cA_z\zbz$ and $\cA_{\bz}\zbz$:
\begin{statement}
\begin{eqnarray}
&&{(\cA_z)}_{j+1,m}\zbz=   \Biggl[\sum_{r=1}^s \sum_{l=0}^j
\left(\begin{array}{c} j\\l\end{array} \right)
\prt^{j-l}{(\cA_z)}_{1,1+(s-r)}\zbz\cF^{(r-l)}_{(s-m+1)}(s;\zbz)\Biggr],\nn\\
\lbl{Azdependent}\\
&&{(\cA_\bz)}_{j+1,m}\zbz=   \Biggl[\sum_{r=1}^s \sum_{l=0}^j
\left(\begin{array}{c} j\\l\end{array} \right)
\prt^{j-l}{(\cA_\bz)}_{1,1+(s-r)}\zbz\cF^{(r-l)}_{(s-m+1)}(s;\zbz)\Biggr].
\nn\\ 
\lbl{Adependence}
\end{eqnarray}
\end{statement}
For all these reasons we now introduce a shorthand notation for the
independent ghosts 
\begin{eqnarray}
\cK_{1,p}\zbz&\equiv& \cK^{(p-1)}\zbz ;\quad p=2,\dots,s
\lbl{notation}
\end{eqnarray}
and we can rewrite the variation \rf{deltaw} only in terms of the very
independent ghosts as:
\begin{eqnarray}
 \delta_\cW  \cK^{(p)}\zbz &=&
\sum_{r=1}^{s-1} \sum_{m=1}^{r}\left(\begin{array}{c}r\\m \end{array}\right)
\cK^{(r)}\zbz\prt^{q} \cK^{(p+m-r)}\zbz\nn\\
&&-\sum_{r=p}^{s-1} \cK^{(r)}\zbz\left(\begin{array}{c}r\\r-p \end{array}\right) \sum_{q=1}^{s-1} \sum_{n=1}^q\left(\begin{array}{c}q\\n \end{array}\right) \prt^{(n+r-s)}\cK^{(n)}\zbz\nn\\
\nn\\
&&+\sum_{r,l=1}^{s-1} \sum_{q=1}^{r}\left(\begin{array}{c}r\\q
\end{array}\right) 
\cK^{(r)}\zbz\prt^{q} \cK^{(l)}\zbz \cF^{((s+q)-(l+r))}_{(s-p)}(s;\zbz)\nn\\
&&-\sum_{r=1}^{s-1}\sum_{q\geq 1}^{r}\cK^{(r)}\prt^q
\biggl\{\sum_{j=1}^{s-1}\sum_{r'=1}^{s-1} \sum_{l=1}^j\left(\begin{array}{c}j\\l\end{array}\right)\prt^{j-l}\cK^{(s-r')}\zbz
\zbz\nn\\
&&\cF^{(r'-l)}_{(s-j)}(s;\zbz)\biggl\} \cF^{(s+q-r)}_{(s-p)}(s;\zbz) .
\lbl{sktotal}
\end{eqnarray}
Suitable transformations for the $\cF^{(q)}_{(p)}(s;\zbz)$ fields
make the algebra to be nilpotent:  in the usual approach for $\cW$-algebras 
\cite{Hull:1992hy} this produces a tower of laws for higher spin
fields, in our approach this information is contained in the
simultaneous use of   Eq \rf{Zvariations0} and  Eqs
\rf{f1}\rf{coefficients}.

An important remark has to be made: if we isolate in
Eq\rf{sktotal} only the first derivative contribution:
\begin{eqnarray}
&& \delta_{\cW,0} \cK^{(p)}\zbz
=\sum_{r=1}^{s-1} r \cK^{(r)}\zbz\prt \cK^{(p-r+1)}\zbz
\lbl{delta0}
\end{eqnarray}
and it is readily checked that  $\delta^2_{\cW,0} =0$.
This property links the present treatment with the one given in
\cite{Bandelloni:1999et,Bandelloni:2000fs,Bandelloni:2000ri},
where a similar ghost transformation has been found within a
symplectic framework, and where finite $\cW$-algebras occur due to a
particular symplectomorphism breaking down mechanism. In the
approach there, to which the reader is referred, the geometrical aspect
plays an essential role and gives a complete representation of each
element  of the algebra in terms of the generating function of
canonical transformations. 
In particular, the holomorphic ghost fields behave as ordinary
tensors (and not as jets), and can be decomposed in non
holomorphic sectors $c^{(p,q)}\zbz$ with well defined $(p,q)$
tensorial characters and geometrical interpretation, transforming
as:
\begin{eqnarray}
\delta_{\cW,0}
c^{(p,q)}\zbz=\sum_{r,s=0...,\ r+s>0}\biggl(rc^{(r,s)}\zbz\prtz
c^{(p-r+1,q-s)}\zbz  \nn \\
+ s\, c^{(r,s)}\zbz\prtbz c^{(p-r,q-s+1)}\zbz\biggr). \lbl{0000a}
\end{eqnarray}
Within this parametrization the contribution to the algebra of the
underived lowest order ghosts $c^{(1,0)}\zbz$, $c^{(0,1)}\zbz$
represents the point displacement (in the $\zbz$ plane) of the
fields, and the full transformation admits breakdown
terms, needed to represent finite $\cW$-algebras
\cite{Bandelloni:2000ri}.
According to \cite{Bandelloni:1988ws}, the ordinary derivative
operator can be in general represented by considering the Fock space of the
monomials and their derivatives, as the anti-commutator of the
B.R.S. operator and the derivative with respect to the first level ghost fields
$c^{(1,0)}\zbz$ and $c^{(0,1)}\zbz$. One thus has
\begin{eqnarray}
\prt=\biggl\{\frac{\prt}{\prt c^{(1,0)}\zbz},\delta
\biggr\};\quad\bprt=\biggl\{\frac{\prt}{\prt c^{(0,1)}\zbz},\delta
\biggr\}.
 \lbl{derivativez}
\end{eqnarray}
In this context we can rederive Eqs\rf{deltaA}, \rf{deltaAb} by
considering the 
$\cA_z\zbz$ and $\cA_\bz\zbz$ as derivatives of the $\cK\zbz$
matrix with respect the previous ghosts fields, namely
\begin{eqnarray}
\cA_z\zbz =\frac{\prt \cK\zbz}{\prt c^{(1,0)}\zbz};\quad
\cA_\bz\zbz =\frac{\prt \cK\zbz}{\prt c^{(0,1)}\zbz}.
\lbl{Aconnect}
\end{eqnarray}
The  $\delta_{\cW,0}$ operator is stable under any change of charts if
 and only if the ghosts behave as true tensors, so, if we want to
 reconstruct the whole algebra Eq\rf{sktotal} on a basis which leaves
 stable this ideal, we have to decompose $\cK^{(r)}\zbz$   in order to
 select its tensorial part. To do this, we introduce the Beltrami
 notation used in \cite{Bandelloni:1999et},
\begin{eqnarray}
&&\prt Z^{(p)}\zbz \equiv \lambda^{(p)}\zbz,\nn\\
&&\bprt Z^{(p)}\zbz \equiv \lambda^{(p)}\zbz\mu(p,\zbz),
\lbl{beltrami0}
\end{eqnarray}
with the usual tricks we can get from the variation Eq \rf{Zvariations0} using
the results coming from the symplectic approach:
\begin{eqnarray}
\lambda^{(j)}\zbz &=&\sum_{l=1}^{s}{\cA_z}_{1,l}\zbz
S^{(j,l)}\zbz
\lbl{lambdaj}\\
 {\lambda^{(j)}\zbz}\mu((j);\zbz)   & =&
 \sum_{l=1}^{s}{\cA_\bz}_{1,l}\zbz S^{(j,l)}\zbz, 
\lbl{muj}
\end{eqnarray}
where we have set
\begin{eqnarray}
S^{(j,l)}\zbz = \sum_{h=0}^{l-2}
\left(\begin{array}{c}l-1\\
h\end{array}\right)\Biggl\{
 \frac{\prt^{l-1-h}{\omega^{-1/s}\zbz}}{\omega^{-1/s}\zbz}\Biggr\}{\prt^h
 \lambda^{(j)}\zbz}.
\lbl{Beltramij}
\end{eqnarray}
With the inversion of the previous Equations we can deduce the
expression of the matrix elements of the first row of the
$\cA_z\zbz$ matrix as function of only $\lambda\zbz's$ and their
$\prt$ derivatives, while the Beltrami's make their appearance in
the  $\cA_\bz\zbz$  matrix.

Then the symplectic  approach allows one to introduce
 ghost tensors $\cC^{(r)}\zbz$ with conformal weight $-r$, $r\geq 1$
and without  loss of generality for $1<j<s$ we can write the matrix
decomposition 
\begin{eqnarray}
\cK\zbz =\sum_{p\geq 0, r\geq 1} \prt^p \cC^{(r)}\zbz \cB^p_r\zbz.
\lbl{Kexpansion}
\end{eqnarray}
In particular, by eq\rf{maximum} we can identify up to a constant factor
\begin{eqnarray}
\cK_{1,s}\zbz \sim \cC^{(s-1)}\zbz, \lbl{maximumKC}
\end{eqnarray}
(see Eq\rf{Kw3} in Appendix C for an explicit example).

Now Equation \rf{Kexpansion} also according to the
results obtained in
\cite{Bandelloni:1999et,Bandelloni:2000fs,Bandelloni:2000ri}, and
Eq\rf{Aconnect}  will imply the matrix decomposition 
\begin{eqnarray}
\cA_\bz\zbz&=& 
\sum_{p\geq 0, r\geq 1} \prt^p \mu_\bz^{(r)}\zbz \cB^p_r\zbz,
\lbl{bilfockK}
\end{eqnarray}
where $ \mu_\bz^{(r)}\zbz$  are the so-called Bilal-Fock-Kogan coefficients
\cite{Bilal:1991wn} :  we can find out this expansion in the
general case. With this result the $\cB\zbz$ matrices can be computed
in order to reconstruct the above expansion \rf{Kexpansion}.

\section{Lagrangian fields Theory models}

The purpose of this Section is to study a dynamical problem in the
$(s-1)$-dimensional space, with local complex coordinates
$Z^{(r)}\zbz$ such that 
the formulation of the dynamics must obey all the
prescriptions given in Statement \rf{covariance}.

For these reasons and considering these local coordinates as scalar
  fields  the symmetry transformations
Eq\rf{Zvariations0} yield an equivalence relation for these
coordinates over each point of the $\zbz$ plane. This means that
the lift of a dynamical model from the  $\zbz$ base space to the
target space with local complex coordinates $Z^{(r)}$ holds its validity if a
$\cW_s$-symmetry is imposed as a dynamical constraint.
Furthermore, if this symmetry is taken as a symmetry principle, the
model can be constructed for any reparametrization of the
$(s-1)$-dimensional target space.

To test this, we construct a toy Action with a scalar field $\cX\zbz$
interacting with the background in a $\cW$-symmetric way. The action reads
\begin{eqnarray}
\Gamma_{S}[\cX,\mu(r)] = \int \cD \cX\zbz \wedge \bar {\cD}\cX\zbz
\lbl{StringAction}
\end{eqnarray}
where:
\begin{eqnarray}
\cD &=& \sum_{r=1}^{s-1} (dz +\mu(r,\zbz)d\bz) \cD_{(r)},
\lbl{derivative}\\
\cD_{(r)}\zbz&\equiv& \frac {\prt
-\mub(r,\zbz)\bprt}{1-\mu(r,\zbz)\mub(r,\zbz)} ,
\lbl{Dr}
\end{eqnarray}
such that$(\cD+\bar {\cD})\cX\zbz$ is globally defined as a 1-form on
the Riemann surface.

\medskip

 The space-time content of this action depends only on the Beltrami
coefficients $\mu(r,\zbz)$. This
implies that the model does not exploit all the  $Z^{(r)}\zbz$
content since only the complex structures of these fields appear
in the Classical limit. This means that the $\lambda^{(r)}\zbz$
sectors, and in particular the fields contained in the $\cA_z\zbz$
matrix, do not appear in the tree Lagrangian.

We must take care of this particularity, since the Quantum corrections
 may excite these degrees of freedom which remain
silent in the Classical level. So a suitable constraint on
them could be required, as we shall see, for the quantization of the
model.

Since the $Z^{(r)}\zbz$ fields infinitesimally transform under a
$\cW_s$-symmetry 
as the BRS  transformations Eq \rf{Zvariations0}, we can
derive:
\begin{eqnarray}
\delta_\cW \mu(r,\zbz)&=&\frac{1-\mu(r,\zbz){\bar\mu}(r,\zbz)}{\prt
Z^{(r)}\zbz}\, {\bar\cD}_{(r)}\zbz\Lambda^{(r)}\zbz,
\lbl{smur}\\
\Biggl[\delta_\cW,\frac{\cD_{(r)}\zbz}{\prt
Z^{(r)}\zbz}\zbz\Biggr]&=&-\sum_{l=1}^{s-1}
\frac{\cD_{(r)}\zbz}{\prt Z^{(r)}\zbz}
\Lambda^{(s)}\zbz\frac{\cD_{(s)}\zbz}{\prt Z^{(s)}\zbz}\nn\\
&+&\frac{\cD_{(r)}\zbz}{\prt Z^{(r)}\zbz}
{\bar\Lambda}^{(s)}\zbz\frac{{\bar\cD}_{(s)}\zbz}{\bprt \bar{Z}^{(s)}\zbz},
\lbl{commD}
\end{eqnarray}
and  the scalar $\cX\zbz$ field transforms under $\cW_s$ as:
\begin{eqnarray}
\delta_\cW \cX\zbz =\sum_{r=1}^{s-1}\Biggl\{\frac{\Lambda^{(r)}\zbz}{\prt
Z^{(r)}\zbz} \cD_{(r)} \cX\zbz
  + \frac{{\bar\Lambda}^{(r)}\zbz}{\bprt {\bar Z}^{(r)}\zbz}{\bar\cD}_{(r)}
\cX\zbz\Biggr\}.
 \lbl{SX}
\end{eqnarray}
  The action  $\Gamma_{S}$  is invariant under the Classical B.R.S
  operator, defined in a functional language as:
 \begin{eqnarray}
  \delta_\cW&=&\int dz\wedge d\bz \Biggl\{\sum_{r=1}^{s-1}
  \Lambda^{(r)}\zbz \frac{\delta}{\delta Z^{(r)}\zbz} +
  \bar{\Lambda}^{(r)}\zbz \frac{\delta}{\delta \bar{Z}^{(r)}\zbz }
  \nn\\
   &+&\Lambda^{(r)}\zbz \frac {\cD_{(r)}\zbz\cX\zbz}{\prt Z^{(r)}\zbz}\frac{\delta}{\delta
   \cX\zbz} +\bar{\Lambda}^{(r)}\zbz
   \frac {{\bar{\cD}}_{(r)}\zbz\cX\zbz}{\bprt
\bar{Z}^{(r)}\zbz} \frac{\delta}{\delta
   \cX\zbz}\nn\\
   &+& \delta_\cW\cK^{(r)}\zbz\frac{\delta}{\delta \cK^{(r)}\zbz} +\delta_\cW\bar{\cK}^{(r)}\zbz\frac{\delta}{\delta \bar{\cK}^{(r)}\zbz}
\Biggr\}.
\lbl{deltaWString}
\end{eqnarray}
Anyhow in order to clarify the reasons why we claim that this model
is defined in a $(s-1)$-dimensional space. Indeed, the equation of
motion of the $\cX\zbz$ field:
\begin{eqnarray}
&&0=\frac{\delta \Gamma_{S}}{\delta \cX\zbz}=
\sum_{l,r=1}^{s-1} \left[
\prt \left(
\frac{\big( 1-\mu(r,\zbz)\mub(l,\zbz) \big)
{\bar\cD}_{(r)}\zbz\cX\zbz}{
1-\mu(l,\zbz)\mub(l,\zbz)} \right)  \right.\nn\\
&& +\ \left.
\bprt \left( \mu(r,\zbz)\,
\frac{\big( 1-\mu(r,\zbz)\mub(l,\zbz) \big)
{\bar\cD}_{(r)}\zbz\cX\zbz}{1-\mu(l,\zbz)\mub(l,\zbz)}
\right) \right] 
\lbl{eqmotstri}\\
&&+\ c.c.\nn
\end{eqnarray}
first gives a solution as a functional of the complex structures
$\mu(r,\zbz)$.
After replacing the latter through \rf{beltrami0}, the scalar field
solution turns out to be a functional 
$\cX\biggl( Z^{(1)}\zbz,\cdots, {Z}^{(s-1)}\zbz;$
$\bar{Z}^{(1)}\zbz,\cdots, \bar{Z}^{(s-1)}\zbz\biggr)$ whose
evolution in terms of the $Z^{(r)}\zbz$ coordinates is fixed by the
equation of motions of the fields $Z^{(r)}\zbz$ themselves,
\begin{eqnarray}
\frac{\delta }{\delta Z^{(r)}\zbz}\int
dz\wedge d\bz
\sum_{l,m=1}^{s-1}\big( 1-\mu(m,\zbz)\mub(l,\zbz)\big) &\times&
\nn\\
\cD_{(m)}\zbz\cX\zbz \cD_{(l)}\zbz\cX\zbz &=& 0, 
\lbl{eqmotionZr}
\end{eqnarray}
in which the expression $\cX\biggl( Z^{(1)}\zbz,\cdots,
{Z}^{(s-1)}\zbz; \bar{Z}^{(1)}\zbz,\cdots, \bar{Z}^{(s-1)}\zbz\biggr)$
obtained in Eq\rf{eqmotstri} must be plugged into.
So the scalar field lives in
a $2(s-1)$-dimensional space  $\biggl( Z^{(1)}\zbz,$ $\cdots,
{Z}^{(s-1)}\zbz;\bar{Z}^{(1)}\zbz,\cdots, \bar{Z}^{(s-1)}\zbz\biggr)$
and its dynamics is fixed by the evolution of the coordinate-fields.

Anyhow, if we want to improve it at the quantum level, we have to
work in the two-dimensional $\zbz$; only in this way we can extend
the $\cW$-algebra as a coordinate symmetry. For this reason,
some external fields are introduced and  a new Classical
Lagrangian is defined to be:
\begin{eqnarray}
&&\Gamma_{Classical}[\cX,\mu(r),\gamma_{Z^{(r)}},\xi_{(r)},\gamma_{\cX}]
= \Gamma_{S}[\cX,\mu(r)] 
+ \int dz\wedge d\bz\Biggl\{\nn\\
&&\lbl{Gammagood}\\
&&\sum_{r=1}^{s-1}\bigg(
\gamma_{Z^{(r)}}\zbz \delta_\cW Z^{(r)}\zbz 
+\ \xi_{(r)}\zbz \delta_\cW \cK^{(r)}\zbz +c.c. \bigg)
+\gamma_{\cX}\zbz \delta_\cW \cX\zbz
\Biggr\},\nn
\end{eqnarray}
 such that at the tree level
\begin{eqnarray}
 \delta_\cW\Gamma_{Classical} =\int dz \wedge d\bz
&\Biggl\{&\sum_{r=1}^{s-1} \biggl[\frac{\delta
\Gamma_{Classical}}{\delta \gamma_{Z^{(r)}}\zbz} \frac{\delta
\Gamma_{Classical}}{\delta Z^{(r)}\zbz} + \frac{\delta
\Gamma_{Classical}}{\delta
\xi^{(r)}\zbz}\frac{\delta\Gamma_{Classical}}{\delta
\cK^{(r)}\zbz}\biggr]
\nn\\
&+& \frac{\delta \Gamma_{Classical}}{\delta \gamma_{\cX}\zbz}
\frac{\delta\Gamma_{Classical}}{\delta
 \cX\zbz}\Biggr\}=0,
\lbl{Classical}
 \end{eqnarray}
and $ \delta_\cW^2 = 0$.

So we search for an Action $\Gamma_Q$ as a Quantum extension of
$\Gamma_{Classical}$ to improve at a Quantum level the symmetry of
Eq\rf{Classical}. So a perturbative calculation program by means,
for example, a Green functions expansion require a renormalization
 which could, in general, spoil the symmetry.
 Anyhow the $\zbz$
locality properties of the theory guarantees that any possible
breakdown is, at the first order of the perturbative extension,
a local term. So we can begin the procedure  with the  Quantum
Action Principle:
\begin{eqnarray}
\delta_\cW \Gamma_Q =\int \Delta_2^1\zbz.
 \lbl{QAP}
\end{eqnarray}
The anomaly  $ \Delta_2^1\zbz$ has already been found in the
literature \cite{Zucchini:1993xm,Abud:1995,Garajeu:1995jn}, with the
classical method of the descent equations.

It can be turned into a well defined expression as
\cite{Garajeu:1995jn,Abud:1995}, $Tr(\cA_\bz\delta_\cW \cA_z - \cK
\prtbz \cA_z)$ which reads upon using the flatness condition
\rf{nullcurvature} and the variation \rf{deltaA}
\begin{eqnarray}
\Delta_2^1\zbz  &=& Tr\bigg[\cA_\bz\zbz \prt \cK\zbz -\cK\zbz
\prt \cA_\bz\zbz
\nn\\
&+&  2 \cK\zbz
 \biggl(\cA_z\zbz \cA_\bz\zbz -\cA_\bz\zbz \cA_z\zbz
\biggr)\bigg] dz\wedge d\bz.
\lbl{wellanomaly}
\end{eqnarray}

The existence of an anomaly, only dependent of the gravitational
content, breaks, at the Quantum level, the equivalence relation of
the $Z^{(r)}\zbz$ coordinates for all points of the $\zbz$ base
space, imposed by Eq \rf{Zfintrans} (and Eq \rf{Zvariations0} in a
B.R.S. framework). This dynamically induced inequivalence brings
new difficulties, for example, in the coordinate ordering
procedure. For this reason the anomaly must be managed in order to
understand what kind of residual symmetry of the Classical one can be
maintained at the Quantum level.

Now enters the compensation of the anomaly in order to restore the
invariance at the quantum level by adding well-defined
counter-terms. First  we point out, as a trailer,  the
crucial step (coming from the anomaly structure) of this
program, and we shall show how to get the on-shell quantum extension of
the model.

Then we re-open the renormalization program with all the dynamical
variables needed for its full completion, and after showing how to
remove  all the trivial anomalies introduced by all these new
degrees of freedom, we prove the consistency of the sketch given
in the trailer.

Note that the previous well-defined anomaly can be rewritten as
 \begin{eqnarray}
 \Delta^1 = 2 \int dz\wedge d\bz\, Tr(\cA_\bz\zbz
\delta_{\cW} \cA_z\zbz) \lbl{wellanomalybis}
 \end{eqnarray}
by dropping out the boundary term $\prtz Tr(\cK\cA_\bz)$.

In order to compensate the anomaly let  us
 introduce an invariant background classical field, namely a
 background $(1,0)$ type connection ${\zer{\cA}}_z\zbz$ 
 \cite{KLS91b}, with
\begin{eqnarray}
\delta_{\cW} {\zer{\cA}}_z\zbz &=& 0,
\lbl{A0}\\
 (w') {\zer{\cA}}_w\wbw &=& \biggl(
\prt_z \Psi^{\frac{1-s}{2}}(z)\biggr) \Psi^{\frac{1-s}{2}}(z)^{-1}
+ \Psi^{\frac{1-s}{2}}(z)
{\zer{\cA}}_z\zbz\Psi^{\frac{1-s}{2}}(z)^{-1}.\nn\\
\lbl{changeconnectA0}
\end{eqnarray}
It is easy to show that
\begin{eqnarray}
&& \delta_{\cW}  \int dz\wedge d\bz\, Tr
\left[\biggl(\cA_z\zbz-{\zer{\cA}}_z\zbz\biggr)
\cA_\bz\zbz\right] = \nn\\
&&\frac{1}{2} \Delta^1 +
 \int dz\wedge d\bz\, Tr
\left[\biggl(\cA_z\zbz-{\zer{\cA}}_z\zbz\biggr)\delta_{\cW}
\cA_\bz\zbz \right], 
\lbl{scounterterm}
\end{eqnarray}
and each term in the right hand side is well defined.

By adding a counter-term in order to modify the
action $\Gamma_S$ into a well-defined one, $\widehat{\Gamma}$,
\begin{eqnarray}
\widehat{\Gamma}_{Tree} =\Gamma_S -2\hbar\int dz\wedge d\bz\ Tr
\left[\biggl(\cA_z\zbz-{\zer{\cA}}_z\zbz\biggr)
\cA_\bz\zbz\right],
 \lbl{newgamma}
\end{eqnarray}
such that at the tree diagrams (but at the first order of $\hbar$) level:
\begin{eqnarray}
\delta_{\cW}\widehat{\Gamma}_{Tree}&=& \delta_{\cW}\biggl(\Gamma_S
-2\hbar\int dz\wedge d\bz\ Tr
\left[\biggl(\cA_z\zbz-{\zer{\cA}}_z\zbz\biggr)
\cA_\bz\zbz\right]\biggr)\nn\\
&&\lbl{trivaria}\\
&=& -\hbar\int \Delta^1_2\zbz dzd\bz +2\hbar
 \int dz\wedge d\bz\ Tr
\left[\biggl(\cA_z\zbz-{\zer{\cA}}_z\zbz\biggr)\delta_\cW
\cA_\bz\zbz \right].\nn
\end{eqnarray}
So the first term in the r.h.s. cancels the local anomaly coming from
an hypothetical Feynman diagrams of the perturbative loop expansion
starting from the action $\Gamma_{Classical}$.
  The quantum improvement thus  leads to an Action
$\widehat{\Gamma}_Q$ such that the anomaly has been shifted to:
\begin{eqnarray}
\delta_\cW\widehat{\Gamma}_Q=2\hbar\int dz\wedge d\bz Tr
\Biggl[\biggl({\cA}_z\zbz-{\zer{\cA}}_z\zbz\biggr)(\delta_\cW
\cA_\bz\zbz)\Biggr],
 \lbl{anomaly}
\end{eqnarray}
such that:
\begin{eqnarray}
\delta_\cW\widehat{\Gamma_Q}\arrowvert_{{{\cA}}_z\zbz={\zer{\cA}}_z\zbz}=0.
\lbl{masshellsymm}
\end{eqnarray}
and the symmetry is restored on the surface
\begin{eqnarray}
{{\cA}}_z\zbz=\zer{\cA}_z\zbz.
 \lbl{constraint}
\end{eqnarray}
As already remarked in the discussion of
Eq\rf{lambdaj} this surface fixes only the derivatives $\prt^n
Z^{(r)}\zbz$, $n\geq 1$ 
and does not affect the complex structures parametrized by the
Beltrami coefficients $\mu(r,\zbz)$
which are coupled, in the tree Action, with the physical degrees of
the scalar field $\cX\zbz$. On the
other hand, this condition removes the arbitrariness on the
definition of the  $a^{(s)}_j\zbz$ coefficients given by \rf{linear}.

This choice looks something similar to a
gauge choice, and identifies, within the structure of the
$Z^{(r)}\zbz$ coordinate-fields, the physical degrees of freedom
from the unphysical ones.
Moreover the renormalization program, needed for the quantum
improvement of the model,  modifies the Action and, more
important, the surface defined by Eq\rf{constraint}, so we have to bring
all under tight control at each step of the quantum improvement.
The presence of the $\zer{\cA}_z\zbz $ fields could introduce
already at the Classical limit arbitrary terms invariant under
$\cW$ transformations.  The invariance under changes of charts (and the
absence of the suitable connection) admits only the term
$\prtbz \zer{\cA}_z\zbz$ which is disconnected .

We shall now discuss this last subject, which is of great
consequence in our treatment, to extend it at any order of a
perturbative expansion.

First of all, we remind that the
 ${{\cA}}_z\zbz$ matrix depends only on the $\lambda^{(r)}\zbz$
fields and their $\prtz$ derivatives, which do not appear in the
tree Lagrangian in Eq\rf{StringAction}, so this constraint do not
alter the dynamics of the $\cX\zbz$ scalar field at the Classical level.

 To clarify this point in a Field Theory language, and to improve it
  at any order of the renormalization perturbative program, 
  we have to put this constraint  working off-shell and perform a
  Faddeev-Popov like
construction working out  Eq \rf{constraint} by means of a
functional integration.

Only the   ${{\cA}}_{(z, (1,l))}\zbz;l=2,\dots, s$ are independent
 (see Eq \rf{Azdependent}) and are calculated by means of Eq
\rf{lambdaj}. We impose:
\begin{eqnarray}
&&\delta_\cW\biggl(\zer{\cA}_{z,1,r+1}\zbz-{{\cA}}_{z,1,r+1}\zbz\biggr)\nn\\
=&&\int
\prod_{r=1}^{s-1}d\beta_{(r)}\exp\left[-\frac{1}{\hbar}\int
dz\wedge d\bz
\beta_{(r)}\zbz\biggl(\zer{\cA}_{z,1,r+1}\zbz-{{\cA}}_{z,1,r+1}\zbz\biggr)
\right],
\lbl{faddpopov}
\end{eqnarray}
to which must be added a further $b_{(r)}\zbz \delta_\cW Z^{(r)}\zbz$
term in order to restore the symmetry, with:
\begin{eqnarray}
\delta_\cW b_{(r)}\zbz &=&-\beta_{(r)}\zbz,
\lbl{sb}\\
\delta_\cW \beta_{(r)}\zbz &=&0.
\lbl{sBeta}
\end{eqnarray}

So the symmetry is preserved on-shell only. Anyhow the quantization
program must  be performed off-shell and we have to check that the
on-shell properties are not wasted.
 In this case, the Quantum extension of the model will treat on the anomaly
cancellation program,
 by means of the counter-term machinery.

For this purpose we introduce already at the tree level the
$\hbar$ order counter-terms  and the Faddeev-Popov  ghost term ; so
the tree Action becomes:
\begin{eqnarray}
\Gamma &=& \Gamma_S + \int dz\wedge d\bz
\biggl\{ \sum_{r=1}^{s-1} \biggl[
\beta_{(r)}\zbz \biggl({{\cA}}_{(z,(1,r+1)}\zbz-\zer{\cA}_{(z,(1,r+1))}\zbz\biggr) \nn\\
&+&  b_{(r)}\zbz\delta_\cW{{\cA}}_{(z,(1,r+1))}\zbz +\gamma_{(r)}\zbz
\delta_\cW Z^{(r)}\zbz + \xi_{(r)}\zbz \delta_\cW \cK^{(r)}\zbz +c.c.\biggr]
\nn\\
&+& \gamma_{\cX}\zbz \delta_\cW \cX\zbz  - 2 \hbar
Tr\Biggl[
\biggl(\cA_z\zbz-{\zer{\cA}}_z\zbz\biggr) \cA_\bz\zbz\Biggr]\nn\\
&-& \theta  Tr\Biggl[
\biggl(\cA_z\zbz-{\zer{\cA}}_z\zbz\biggr) \delta_\cW\cA_\bz\zbz\Biggr
]\Biggr\} \nn\\
\nn\\
&\equiv& \Gamma_0+\hbar \Gamma_{1,Tree} +\theta \Gamma_\theta
\lbl{gammatotal}
\end{eqnarray}
where we have introduced $\theta$ as a para-fermionic coordinate such
that:
\begin{eqnarray}
\delta_\cW \theta =\hbar \lbl{deltatheta}
\end{eqnarray}
in order to  control the anomalous term defined in Eq\rf{anomaly}.
To perform it, and to show the completeness of the quantum perturbative
expansion, we must switch to functional techniques and introduce by 
Legendre transformation the connected Green gnerating functional
\begin{eqnarray}
\cZ_c[\cJ_{\cX},\cJ_{Z^{(r)}},\cJ_{\cK^{(r)}},...]
&=&\Gamma+\int dz\wedge d\bz \bigg[ \cJ_{\cX}\zbz \cX\zbz
\nn\\
&& \hskip -3cm \left.
+ \sum_{r=1}^{s-1} \cJ_{Z^{(r)}}\zbz Z^{(r)}\zbz
+\cJ_{\cK^{(r)}}\zbz \cK^{(r)}\zbz +c.c.\bigg]
\right|_{\small{\left[
\begin{array}{c}
\cJ_{\cX}=-\frac{\delta \Gamma}{\delta  \cX}\\
\cJ_{\cK^{(r)}}=-\frac{\delta \Gamma}{\delta  \cK^{(r)}}\\
\cJ_{Z^{(r)}}=-\frac{\delta \Gamma}{\delta  Z^{(r)}}\\
\end{array}\right.}}
\lbl{Zconnected}
\end{eqnarray}
and we can construct the Green generating functional:
\begin{eqnarray}
\cZ[\cJ_{\cX},\gamma_{\cX},\zer{\cA}_z{\!},\cJ_{\cK^{(r)}},\xi_{(r)},
\gamma_{Z^{(r)}},c.c] =\int d\cX \prod_{r=1}^{s-1}
dZ^{(r)} db_{(r)} d\beta_{(r)}d\cK^{(r)} \exp[-\frac{1}{\hbar}
\cZ_c].
\lbl{Ztotal}
\end{eqnarray}

In this description the interpretation of the coordinates as quantum
fields is particularly suitable to manage the several unexpected
events in the quantization of the model.  Anyhow the role of
coordinates, understood as passive entity for events description,
has to be widely amended at the quantum  level, while at the
classical approximation has to maintain all its geometrical
properties.

The B.R.S operator acting on $\cZ_c$ writes
\begin{eqnarray}
\delta_{\cZ_c} &=& \int dz\wedge d\bz \Biggl[
\cJ_{\cX}\zbz\frac{\delta}{\delta \gamma_{\cX}\zbz} +
\sum_{r=1}^{s-1}\biggl(\cJ_{Z^{(r)}}\zbz\frac{\delta}{\delta
\gamma_{Z^{(r)}}\zbz}\nn\\
&&-\ \beta_{(r)}\zbz\frac{\delta}{\delta
b_{(r)}\zbz}+\cJ_{\cK^{(r)}}\frac{\delta}{\delta
\cK^{(r)}\zbz}+c.c. \biggr)+\hbar \frac{\delta}{\delta
\theta}\Biggr].
\lbl{brstotal}
\end{eqnarray}

Our program is to show now that at every order of the perturbative
theory we can define a quantum improvement of both the Action and the
B.R.S. operator such that:
\begin{eqnarray}
\delta_{\cZ_c} \cZ_c=0.
 \lbl{symmetry}
\end{eqnarray}

For this reason, since we principally look for an extension of the
classical Action, we introduce the linearized operator acting on
$\Gamma_0$, which will be useful in the sequel:
\begin{eqnarray}
\delta_{L_0} &=& \int dz \wedge d\bz
\biggl\{ \frac{1}{2} \frac{\delta
\Gamma_0}{\delta \gamma_{\cX}\zbz} \frac{\delta}{\delta
 \cX\zbz}
+ \frac{1}{2}\frac{\delta\Gamma_0}{\delta
 \cX\zbz}\frac{\delta }{\delta \gamma_{\cX}\zbz}\nn\\
&&+\ \sum_{r=1}^{s-1} \biggl[\frac{1}{2}\frac{\delta
\Gamma_0}{\delta \gamma_{Z^{(r)}}\zbz} \frac{\delta }{\delta
Z^{(r)}\zbz} +\frac{1}{2}
 \frac{\delta
\Gamma_0}{\delta Z^{(r)}\zbz}\frac{\delta }{\delta
\gamma_{Z^{(r)}}\zbz}\nn\\
& +& \frac{1}{2} \frac{\delta \Gamma_0}{\delta
\xi^{(r)}\zbz}\frac{\delta}{\delta \cK^{(r)}\zbz}
+\frac{1}{2}\frac{\delta\Gamma_0}{\delta
\cK^{(r)}\zbz}\frac{\delta }{\delta \xi^{(r)}\zbz}
\nn\\
 & -&\beta_{(r)}\zbz \frac{\delta}{\delta b_{(r)}
\zbz}+c.c\biggr]
 +\hbar \frac{\delta \Gamma_0}{\delta \theta}
 \Biggr\}
=0,\nn\\
\delta_{L_0}^2 &=&0, \qquad
\delta_{L_0} = \delta_{L_0}^0+\hbar\delta_{L_0}^1.
\lbl{deltalinear}
\end{eqnarray}
 If we consider the Classical level at  $\hbar=0$, we have:
\begin{eqnarray}
\delta_{L_0}\Gamma_0=0.
\lbl{delta0cl}
\end{eqnarray}
The Feynman graph perturbative calculations require a local
counter-term adjustment procedure which in general spoils the Classical
symmetry and modifies the Classical Action.  But at the  same first
order of these corrections, we have to add the tree contributions
previously introduced. More precisely  these tree contributions
modify the B.R.S symmetry equation by

\begin{eqnarray}
\delta_{L_0}\Gamma_{1,Tree}= \frac{1}{2}\int dz\wedge d\bz
\sum_{r=1}^{s-1}\biggl[\frac{\delta \Gamma_{1,Tree}}{\delta
Z^{(r)}\zbz}\frac{\delta\Gamma_0 }{\delta
 \gamma_{Z^{(r)}}\zbz}
 +c.c \biggr] . \nn\\
 \lbl{deltah1}
 \end{eqnarray}
 Moreover the $\theta$ dependent term of the B.R.S. operator (we
 shall use this trick only here, in order to keep the stability of he
calculation) introduce the local term which compensates the anomaly
 \rf{anomaly} since:
\begin{eqnarray}
\delta_{L_0}^1\Gamma_0=-\hbar \Gamma_\theta \lbl{deltaltheta}
 \end{eqnarray} 
So if the anomaly of this new Action is still the one
 of Eq \rf{anomaly}, (i.e. no new anomaly occurs, due to the
 introduction of plenty of such new fields) the counter-term will
 perform its duty and the anomaly will disappear. Thus if no nasty
 surprise appears, as we shall see in the following, the
 renormalization can be  extended to all orders.

\medskip

The $\delta({{\cA}}_z\zbz-{\zer{\cA}}_z\zbz)$ constraint is found to be
 the equation of motion for the $\beta\zbz$ field:
\begin{eqnarray}
0=\frac{\delta \Gamma}{\delta \beta_{(r)}\zbz}=
{{\cA}}_{z,1,r+1}\zbz-{\zer{\cA}}_{z,1,r+1}\zbz.
\lbl{betaequation}
\end{eqnarray}
Note that the constraint Eq\rf{betaequation} is not B.R.S.
invariant since
\begin{eqnarray}
\delta_{L_0} \frac{\delta  \Gamma_0}{\delta \beta_{(r)} \zbz} &=&
\frac{\delta  \Gamma_0}{\delta b_{(r)}\zbz}
- \frac{1}{2} \int d\zp\wedge d\bzp\sum_{m=1}^{s-1}
\frac{\delta \cA_{z,1,r+1}\zbz}{\delta
Z^{(m)}\zbzp}\times \nn\\
&& \hskip 4cm
\frac{\delta  \Gamma_0}{\delta \gamma_{(m)}\zbzp}.
 \lbl{Symmbetaeq}
\end{eqnarray}

So $\beta_{(r)}\zbz$-dependent counter-terms would a priori modify this
equation of motion. In Appendix B, the analysis of the anomalies depending
on the external fields is performed. It is shown there that they become trivial
and  can be compensated by means of local counter-terms. In particular
$\beta_{(r)}\zbz$-dependent anomalies can be compensated by means of
counter-terms with no $\beta_{(r)}\zbz$-dependence  and so the
condition Eq\rf{constraint}, expressed through the 
$\beta_{(r)}\zbz$ mass-shell condition Eq\rf{betaequation}  is
not affected by the renormalization.

Thus the renormalization of external field independent anomalies
can be carried out and it can be easily realized that the
analysis and the results already encountered in Eq\rf{wellanomaly}
can be repeated again.

The main feature of the model is that the  well defined procedure
for counterterms  introduces  pure gravitational
terms of order $\hbar$ which define a local 
propagators $<Z^{(r)}\zbz Z^{(s)}\zbzp>$ also of order $\hbar$ whose
the form depends on the rank of the $\cW$-symmetry. 

We shall investigate this fact, with detailed calculations  in
Appendix C, by means of Ward identities, in the
$\cW_2$ case (where the
 $\ZBZ$ space has the same dimension of the background $\zbz$) and in
the $\cW_3$ one, where the
four-dimensional $Z^{(r)}\zbz$ space is of
particular interest.

The Ward identities can be derived, as well known, from the B.R.S.
equation Eq\rf{symmetry} (or  equivalent).
 Anyhow the inequivalent behavior of the $\cK^{(r)}\zbz$ (jet-) ghosts
under change of charts, produces a complete lack of homogeneity
between the Ward identities obtained from functional
differentiation with respect these ghosts. For this reason we turn
to the $\cC^{(r)}\zbz$ ghost fields, whose good tensorial behavior,
 is suitable for
handling  the identities. Accordingly, the decomposition Eq\rf{Kexpansion} must
be at our disposal and this is the reason why  we have to fix the order of
the $\cW$-algebra in order to push further ahead the calculation.

Indeed, the symmetry conservation and the presence of the local
$\hbar$ pure-gravitational counter-term induce(as we shall see in
detail in the third Appendix), in the $\cW(2)$ case, the local two
point coordinate function:

\begin{eqnarray}
\left.
\frac{\delta^2\Gamma_Q}{\delta Z'\zbzp\delta
Z\zbz} \right|_{\left\{\begin{array}{c}\cA_z=\zer{\cA}_z\\
 <\cX>=0\end{array}\right.} &\equiv& <  Z'\zbzp Z\zbz >
\lbl{propagatorw2}\\
&=&\left.
\frac{\hbar}{\prt Z\zbz}\biggl[\bprt
-\mu\zbz\prt\biggr]\frac{\delta\cT\zbz}{\delta Z'\zbzp}
\right|_{\left\{\begin{array}{c}\cA_z=\zer{\cA}_z
 \end{array}\right.}\nn
\end{eqnarray}
(where $\cT\zbz$ is a projective connection
\cite{A.R.Forsyth:Forsyth,DiFrancesco:1991qr,Gieres:1993sd}
which appears as an entry in the $\cA\zbz$ matrix)
 while in the $\cW(3)$ case, a rather elaborated calculation
allows to reach  well defined results (in  Appendix C the
$<Z^{(1)}\zbz Z^{(1)}\zbzp>$ two-point function is explicitly given).

It is a matter of straightforward calculation that  the $\zbz$ locality of
the model assure a noncommutative imprinting for all Green
functions we can calculate by means of Ward identities.
In particular in the $\cW(2)$ case we easily derive:
\begin{eqnarray}
&&\left.
\Biggl[\frac{\delta}{\delta Z'\zbzp},\frac{\delta}{\delta
Z\zbz}\Biggr]\Gamma_Q \right|_{\left\{\begin{array}{c}\cA= \zer{A}_z\\
 <\cX>=0\end{array}\right.}   \equiv <\bigg[  Z'\zbzp, Z\zbz \bigg]
 >\nn\\
 &=& \Biggl(\frac{\hbar}{\prt
Z\zbz}\biggl[\bprt -\mu\zbz\prt\biggr]\frac{\delta\cT\zbz}{\delta
Z'\zbzp}\nn\\
&&\left. -\ \frac{\hbar}{\prt Z'\zbzp}\biggl[\bprt'
-\mu'\zbzp\prt'\biggr]\frac{\delta\cT'\zbzp}{\delta Z\zbz}
\Biggr) \right|_{\left\{\begin{array}{c}\cA_z=\zer{\cA}_z 
 \end{array}\right.}.
 \lbl{commutatorw2}
\end{eqnarray}

We emphasize that this commutation rule highly depends on the
$\zbz$ background directions. In particular, along the real axis
$z=\bz$ and $z'=\bz'$ with $z\neq z'$ the commutator vanishes.

In higher dimensions the order of the $\cW$-algebra must be increased,
and the results acquire a rather elaborated complexity.
Moreover the examples  clarify the dichotomic role of
$Z^{(r)}\zbz$ as fields and local complex coordinates. In fact when they are
upgraded to quantum fields, non trivial commutation relations will
qualify them as noncommutative coordinates, so the present treatment
acquires some new perspectives.

\section{Conclusions}

The lesson we can derive from our calculations is that the $\cW$-symmetry
already hides many unknown aspects, and its properties
are not completely used within a physical scenario.

In this context we have constructed a Lagrangian Field Theory
model over a complex two-dimensional Riemannian background
manifold where the ``true" complex coordinates are defined \`a la
Laguerre-Forsyth.

Using a scalar field as a probe for the coordinates, we show that the
interaction of
this field with the background, introduces, in the quantum
extension of the model, non trivial two-point functions in the
coordinate fields, which generate an induced quantum gravity over
the Riemann surface. The fields become on-shell "coordinates"  when the
scalar field  equation  is solved, and the space which arises turns
out, by construction, to be noncommutative.

The model can be extended to all orders in the $\hbar$ perturbative
expansion, but
its on-shell peculiarity greatly limits its value.

The construction presented in the paper for the particular case of
Riemann surface raises the issue whether this way of generating
noncommutative space-time coordinates from a quantum field approach
may be (or not) also generalizable to higher complex dimensions, one may
think of the two-complex dimensional (or four-real dimensional) case.
The field coordinates do not commute due to quantum corrections and one
may ask whether this property can be embedded in an algebraic way which
might generalize the many products (Moyal, Kontsevitch..) found in
the literature.

\indent

\noindent
Acknowledgments: We thank Alberto Blasi and Nicola Maggiore for a lot of patience in discussions.

\appendix
\section{Higher orders derivatives of the solutions~$f_i$ }

 Equations \rf{solutionsf} say that the $s$-th
order of derivative of $f_i\zbz$ can be expressed in terms of the
lowest order ones $\prt^j f_i\z\quad j=1\cdots s-1$. Furthermore if
we derive at all orders the previous  equations,
 we can express all the derivatives of $f_i\zbz$ with order
  $l\geq s+1$ with the ones with order $j=1\cdots s-1$, which thus form
a basis of the all arbitrary functions which can be expressed in terms
of $f_i\zbz$ and their derivatives.
So we, after some work, we deduce:
\begin{statement}
\begin{eqnarray}
 \prt^{m+s}f_i\zbz & =& \sum_{k=1}^s\cF_{(k)}^{(-m)}(s;\zbz)
 \prt^{s-k}f_i\zbz ,
\lbl{statement0a}
\end{eqnarray}
where   for $m\geq 0$
\begin{eqnarray}
&&\cF_{(k)}^{(-m)}(s;\zbz)=\nn\\
&&\left.
\sum_{i=1}^{\bar{i}} {(-1)}^i
\Biggl[\prod_{l_i,j_i}\left(\begin{array}{c}
m-\sum_{r=1}^{i-1}(l_r +j_r)\\l_i\end{array}\right)\prt^{l_i}
a^{(s)}_{j_i}\zbz\Biggr] \right|_{\left\{
\begin{array}{c}
 1\leq j_i\leq s\\
0\leq l_i\leq  m-\sum_{r=1}^{i-1}(l_r +j_r)\\
k+m=\sum_{r=1}^{\bar{i}}(l_r +j_r) \end{array}\right.} \lbl{f1}
\end{eqnarray}
while we can trivially extend the previous formula
 for $-1\geq m\geq -s$ defining:
\begin{eqnarray}
\cF_{(k)}^{(-m)}(s;\zbz)=\delta_k^{-m} .
\lbl{f2}
\end{eqnarray}
\end{statement}
We shall call this the "trivial extension" of Eq \rf{statement0a}.
So our definition of $\cF_{(k)}^{(-m)}(s;\zbz) $ will gather
together both Eqs \rf{f1} and\rf{f2}.

\section{External Field dependence of the  anomalies}

 The Q.A.P gives, after the introduction of the $\hbar$
 counterterm:
 \begin{eqnarray}
 \delta_{L_0}\Gamma&=&\int dz\wedge d\bz \Delta\zbz =\int dz\wedge
 d\bz\Biggl[\Delta_0\zbz+\sum_{r=1}^{s-1}\biggl( b_{(r)}\zbz
 \Delta_{b_{(r)}}\zbz\nn\\
 &+&\beta_{(r)}\zbz \Delta_{\beta_{(r)}}\zbz
  +\xi_{(r)}\zbz \Delta_{\xi_{(r)}}\zbz
  +\gamma_{\cX}\zbz\Delta_{\gamma_{\cX}}\zbz\nn\\&+&\gamma_{Z^{(r)}}\zbz\Delta_{\gamma_{Z^{(r)}}}\zbz
 +c.c\biggr)\Biggr]
 \lbl{Deltadecomp}
  \end{eqnarray}
where all the $\Delta's$ terms can be only functions of the fields
$Z^{(r)}\zbz$ ,$\cK^{(r)}\zbz$, $\cX\zbz$ and their derivatives.

It is possible to show, following the tricks contained in
\cite{Bandelloni:1988ws}, that the cohomology space of
$\delta_{L_0}$ in the local functions depending on $Z^{(r)}\zbz$,
$\cK^{(r)}\zbz$, $\cX\zbz$ and their derivatives, and with
Faddeev-Popov charge less than $s-1$ is empty. First one gets
\begin{eqnarray}
&&\delta_\cW\Delta_{\gamma_{(r)}}\zbz\equiv\cD^{(r)}_{(2)}\zbz\Delta\nn\\
&&=\frac{1}{2}\int d\zp\wedge d\bzp \sum_{m=1}^{s-1}\Biggl[
\frac{\delta \delta_\cW Z^{(r)}\zbz}
{\delta Z^{(m)}\zbzp}\Delta_{\gamma_{(m)}}\zbzp +\frac{\delta \delta_\cW Z^{(r)}\zbz}{\delta \cK^{(m)}\zbzp}\Delta_{\xi_{(m)}}\zbzp\Biggr]\nn\\
\lbl{sdeltabg}
\\
&&\delta_\cW \Delta_{\xi_{(r)}\zbz}\zbz\equiv\cD^{(r)}_{(3)}\zbz\Delta \nn\\
&&=\frac{1}{2}\int d\zp\wedge d\bzp \sum_{m=1}^{s-1}\Biggl[
\frac{\delta \delta_\cW \cK^{(r)}\zbz}{\delta
Z^{(m)}\zbzp}\Delta_{\gamma_{(m)}}\zbzp
+\frac{\delta \delta_\cW \cK^{(r)}\zbz}{\delta \cK^{(m)}\zbzp}\Delta_{\xi_{(m)}}\zbzp\Biggr]\nn\\
\nn\\
&& \lbl{sdeltaxi}
\end{eqnarray}
Applying again the B.R.S operator we reach a two-dimensional
system which leads to the trivial solutions:
\begin{eqnarray}
\Delta_{\xi_{(r)}}\zbz &=&\delta_{L_0} \frac{\delta
\widehat{\Gamma}_1}{\delta \xi_{(r)}\zbz} + \cD^{(r)}_{(3)}\zbz
\widehat{\Gamma}_1 ,
\lbl{cobordxi}\\
\Delta_{\gamma_{(r)}}\zbz &=& -\delta_{L_0} \frac{\delta
\widehat{\Gamma}_2}{\delta \gamma_{(r)}\zbz}
+\cD^{(r)}_{(2)}\zbz\widehat{\Gamma}_2, 
\lbl{cobordgamma}
\end{eqnarray}
where the $\widehat{\Gamma}$'s are unknown counterterms. Second, one has
\begin{eqnarray}
&&\delta_\cW\Delta_{b_{(r)}}\zbz \equiv \cD_{(1)}^{(r)}\zbz\Delta\nn\\
&&=\frac{1}{2}\int d\zp\wedge d\bzp \sum_{m=1}^{s-1}\Biggl[
\frac{\delta \delta_\cW \cA_{(1,1+r)}\zbz}{\delta Z^{(m)}\zbzp}\Delta_{
\gamma_{(m)}}\zbzp +\frac{\delta \delta_\cW \cA_{(1,1+r)}\zbz}{\delta
\cK^{(m)}\zbzp}
\Delta_{\xi_{(m)}}\zbzp\Biggr]\nn\\
\lbl{Sdeltab}
\end{eqnarray}
whose solution reads
\begin{eqnarray}
\Delta_{b_{(r)}}\zbz=\delta_{L_0} \frac{\delta\widehat{\Gamma}_3}{\delta
b_{(r)}\zbz}  + \cD^{(r)}_{(1)}\zbz
\widehat{\Gamma}_3 .
\lbl{deltab}
\end{eqnarray}
Finally  we get:
\begin{eqnarray}
&&\hs{-10}\delta_\cW \Delta_{\beta_{(r)}}\zbz = \Delta_{b_{(r)}}\zbz-
\Biggl[ \frac{1}{2}\int d\zp\wedge d\bzp\sum_{m=1}^{s-1}
\frac{\delta \cA_{(z,(1,r+1))}\zbz}{\delta
Z^{(m)}\zbzp}\Delta_{\gamma_{(m)}}\zbzp \Biggr]\nn\\
&=&\delta_{L_0}\frac{\delta}{\delta b_{(r)}\zbz}
\widehat{\Gamma}_3 +
\cD^{(r)}_{(1)}\zbz \widehat{\Gamma}_3\nn\\
&-& \Biggl[ \frac{1}{2}\int d\zp\wedge d\bzp\sum_{m=1}^{s-1}
\frac{\delta \cA_{(z,(1,r+1))}\zbz}{\delta Z^{(m)}\zbzp}\Biggl(
-\delta_{L_0} \frac{\delta \widehat{\Gamma}_2}{\delta
\gamma_{(m)}\zbzp} +\cD^{(m)}_{(2)}\zbzp\widehat{\Gamma}_2\Biggr)
 \Biggr]\nn\\
 \lbl{sdeltabeta}
\end{eqnarray}
with general solution 
 \begin{eqnarray}
 \Delta_{\beta_{(r)}}\zbz=
\delta_{L_0} \frac{\delta}{\delta
\beta_{(r)}\zbz}\widehat{\Gamma}_4
  - \cD_0^{(r)}\zbz {\widehat\Gamma}_4.
  \end{eqnarray}
It is readily seen that the counter-term ${\widehat\Gamma}_4= \int
 dz\wedge d\bz 
b_{(r)}\zbz \Delta_{\beta_{(r)}}\zbz $ removes the anomaly
without any modification of the $\beta_{(r)}\zbz$ sector of the
Lagrangian, and then with no change of the $\beta_{(r)}\zbz$
equations of motions Eq\rf{betaequation}. This counter-term also
benefits the cancellation of $b_{(r)}\zbz$ dependent anomalies (see
Eq\rf{deltab}) by means the use of  Eq\rf{sdeltabeta}
consistency condition.

\section{Ward identities}

As said before, the Ward identities can be derived from the BRS
variation Eq \rf{symmetry} by means of functional differentiation
with respect the ghosts fields. Anyhow the different covariance
behaviors under changes of charts of the $\cK^{(r)}\zbz$ ghosts
produce, if we adopt these ghosts for our procedure, various
Ward identities which cannot be compared among themselves. This is
mainly for this reason that the $\cC^{(r)}\zbz$ ghosts are preferred
and fit well to our requirements.
However the decomposition Eq\rf{Kexpansion} is too difficult to be
managed in a general case and so we limit our analysis to the two
lowest orders of $\cW$-algebras for the sake of simplicity.

\subsect{The $\cW_2$ case}

In this case the $\cK\zbz$ ghost  is at the top and is a $(1,0)$-type
tensor, and the $Z\zbz$ transformation law together with the
Eq\rf{derivativez}  give the relations:
\begin{eqnarray}
\delta_\cW Z\zbz &=& \cC\zbz\prt Z\zbz, \nn\\
\bigg(\bprt -\mu_\bz^z\zbz \prt\bigg) Z\zbz &=& 0, \qquad
\mu_\bz^z\zbz\ =\ \frac{\prt\cC\zbz}{\prt c^{(0,1)}\zbz} 
\lbl{sZ}
\end{eqnarray}
The $\cA_z\zbz$ and $\cA_{\bz}\zbz$  fields satisfying Eq\rf{nullcurvature} can be grouped into:
\begin{eqnarray}
\cA\zbz &\equiv& \cA_z\zbz dz+\cA_{\bz}\zbz d\bz
= \left(\begin{array}{cc}0&1\\ -\frac{1}{2}\cT\zbz &0
\end{array}\right) dz \nn\\
&& +\
\left(\begin{array}{cc}-\frac{1}{2}\prt\mu_\bz^z\zbz&\mu_\bz^z\zbz\\-\frac{1}{2}\mu_\bz^z\zbz
\cT\zbz-\frac{1}{2}\prt^2\mu_\bz^z\zbz
&\frac{1}{2}\prt\mu_\bz^z\zbz \end{array}\right) d\bz.
\end{eqnarray}
The  ``well-defined" anomaly can be deduced from Eq
 \rf{wellanomaly} and writes as \cite{Garajeu:1995jn}:
\begin{eqnarray}
\sS \Gamma_S =\int\Delta^1_2\zbz&=&\int\frac{1}{2}\biggl[\mu_\bz^z
L_3 \cC\zbz-\cC\zbz L_3\mu_\bz^z\zbz \biggr] dz\wedge d\bz,
\lbl{qapw2}
\end{eqnarray}
where we indicate  the third order Bol operator
\cite{G.Bol:bol,Gieres:1993sd} $L_3$ and the projective
connection $\cT\zbz$
\begin{eqnarray}
L_3&=&\prt^3+2\cT\zbz\prt +\prt\cT\zbz, \\
\cT\zbz&=& \prt^2\ln \prt Z\zbz  -\frac{1}{2}{(\prt \ln \prt Z\zbz )}^2\equiv\biggl \{Z\zbz,z\biggr\}\nn \\
&=& \frac{\prt^3 Z\zbz}{\prt
Z\zbz}-\frac{3}{2}{\biggl(\frac{\prt^2 Z\zbz}{\prt Z\zbz}\biggr)}^2 .
\lbl{l3}
\end{eqnarray}
If we introduce a $\cW$-invariant background connection
$\zer{\cA}_z\zbz$, we can build a $\cW$ invariant projective
connection $\zer{\cT}\zbz$:
\begin{eqnarray}
\delta_\cW \zer{ Z}\zbz = \delta_\cW \zer{\cT}\zbz = 0.
\lbl{zerT}
\end{eqnarray}
So our Eq\rf{scounterterm} as a trick yields a partial
compensation of the anomaly by means of well defined counterterms:
\begin{eqnarray}
\delta_\cW\int dz\wedge d\bz \mu_\bz \biggl(\cT\zbz-\zer{\cT}\zbz\biggr)
&=&\int dz\wedge d\bz \biggl[\frac{1}{2}\biggl(\mu_\bz^z\zbz L_3\cC\zbz -\cC\zbz L_3\mu_\bz^z \zbz\biggr)\nn\\
&+&(\delta_\cW\mu_\bz^z\zbz)\biggl(\cT\zbz-\zer{\cT}\zbz\biggr)\biggr].
\lbl{cow2}
\end{eqnarray}
Now, if we define a new $\Gamma_{Tree}$ action:
\begin{eqnarray}
\Gamma_{Tree} =\Gamma_S -2 \hbar\int dz\wedge
d\bz\ \mu_\bz^z\zbz\biggl(\cT\zbz-\zer{\cT}\zbz\biggr),
\lbl{gammatree}
\end{eqnarray}
our procedure  allows to define a new quantum action $\Gamma_Q$, as in
Eq\rf{newgamma}, which gives a new anomalous Ward identity (at zero external
Classical currents):
\begin{eqnarray}
\Biggl(\prt Z\zbz \frac{\delta}{\delta Z\zbz} +\frac{(\prt
-\mub_z^\bz\zbz\bprt)\cX\zbz}{1-\mu_\bz^z\zbz\mu_z^\bz\zbz}
\frac{\delta}{\delta \cX\zbz}\Biggr)\Gamma_Q\nn\\
=-\hbar \biggl[\bprt
-\mu\zbz\prt\biggr]\biggl(\cT\zbz -\zer{\cT}\zbz\biggr).
\lbl{wardw2}
\end{eqnarray}
The $Z\zbz$ tadpole term is zero at zero string v.e.v. and at
$<\cT\zbz>= \zer{\cT}\zbz$:
\begin{eqnarray}
\left.
\frac{\delta \Gamma_Q}{\delta Z\zbz}
\right|_{\left\{\begin{array}{c}\cT=\zer{\cT}\\ 
 <\cX>=0\end{array}\right.}
 = \left. \frac{-\hbar}{\prt Z\zbz} [\bprt -\mu\zbz\prt]\biggl(\cT\zbz
-\zer{\cT}\zbz\biggr) \right|_{\left\{\begin{array}{c}\cT=\zer{\cT}\\
 \end{array}
 \right.}=0,
 \lbl{tadpolew2}
\end{eqnarray}
but a non-trivial $<Z\zbz Z\zbzp>$ two-point function can be derived
at the first order in $\hbar$. The results shown in Eq\rf{propagatorw2}
and\rf{commutatorw2} can be verified.
Note that if $\frac{\delta\cT\zbz}{\delta Z\zbzp}$ is an arbitrary
function of $Z(z-z',\bar{z}-\bar{z}')$ then the propagator and the
commutator are zero.

Higher order Green functions can be derived iterating the
functional derivation process, they appears at the $\hbar$ order
and are non-zero.

\subsect{The $\cW_3$ case}

The correspondence law between the ghosts $\cK^{(s)}\zbz$ and
$\cC^{(s)}\zbz$ for the $\cW_3$-algebra can be found in several papers
\cite{Abud:1995,Bilal:1991wn}:
\begin{eqnarray}
&&\cK_{1,1}\zbz =  \frac{8}{3}\frac{a_2^{(3)}\zbz }{2}\Ch\zbz- \partial\cC\zbz  +\frac{1}{3} \partial^2\Ch\zbz ,    \nn\\
&&\cK_{1,2}\zbz=\cC\zbz -\prt \Ch\zbz, \lbl{Kw3}\\
&& \cK_{1,3}\zbz= 2\Ch\zbz.\nn
\end{eqnarray}
And the $Z^{(r)}\zbz r=1,2$ coordinate transformations can be
written\cite{Bandelloni:2000ri}:
\begin{eqnarray}
\delta_\cW Z^{(1)}\zbz &=& \cC^{(1)}\zbz\partial Z^{(1)}\zbz +
 2\Ch\zbz \partial^2 Z^{(1)}\zbz \nn\\
&& -\ \partial\Ch\zbz \partial Z^{(1)}\zbz - \frac{4}{3} \Ch\zbz
\partial Z^{(1)}\zbz \partial\ln w\zbz,
\lbl{sZ1}\\
\delta_\cW Z^{(2)} \zbz&=& \cC^{(1)}\zbz\partial Z^{(2)}\zbz +
 2\Ch\zbz \partial^2 Z^{(2)}\zbz \nn\\
&& -\ \partial \Ch\zbz \partial Z^{(2)} \zbz- \frac{4}{3} \Ch\zbz
\partial Z^{(2)}\zbz \partial\ln w\zbz, 
\lbl{sZ2}
\end{eqnarray}
where we have indicated the derterminants \cite{A.R.Forsyth:Forsyth}
\begin{eqnarray}
w \zbz = \left| \begin{array}{cc}
\prt Z^{(1)}\zbz  & \prt Z^{(2)}\zbz \\
\prt^2 Z^{(1)} \zbz & \prt^2 Z^{(2)} \zbz
\end{array} \right|, \quad
v\zbz = \left| \begin{array}{cc}
\prt^2 Z^{(1)} \zbz & \prt^2 Z^{(2)}\zbz \\
\prt^3 Z^{(1)} \zbz & \prt^3 Z^{(2)}\zbz
\end{array} \right|.
\lbl{determinants}
\end{eqnarray}
The arguments already given in Eq\rf{derivativez} allow to write
the $\bprt$ derivatives of the coordinates as:
\begin{eqnarray}
\bprt Z^{(1)}\zbz &=& \mu_\bz^z\zbz\partial Z^{(1)}\zbz +
 2\mu^{(2)}_\bz\zbz \partial^2 Z^{(1)}\zbz \nn\\&-&
\partial\mu^{(2)}_\bz\zbz\partial Z^{(1)}\zbz - \frac{4}{3}
\mu^{(2)}_\bz\zbz\zbz
\partial Z^{(1)}\zbz \partial\ln w\zbz\nn\\
\bprt Z^{(2)} \zbz&=& \mu^z_\bz\zbz\zbz\partial Z^{(2)}\zbz +
 2\mu^{(2)}_\bz\zbz\zbz \partial^2 Z^{(2)}\zbz \nn\\
& -& \partial \mu^{(2)}_\bz\zbz\zbz \partial Z^{(2)} \zbz-
\frac{4}{3} \mu^{(2)}_\bz\zbz\zbz \partial Z^{(2)}\zbz \partial\ln
w\zbz
     \nn\\
     \mu^{(2)}_\bz\zbz &=&\frac{\prt\cC^{(2)}\zbz}{\prt
     c^{(0,1)}\zbz}\nn\\
     \mu_\bz^z\zbz &=&\frac{\prt\cC^{(1)}\zbz}{\prt
     c^{(0,1)}\zbz},
     \lbl{prtbZw3}
\end{eqnarray}
so the Beltrami multipliers can be written as:
\begin{eqnarray}
\mu_\bz^z(1,\zbz) \equiv \frac{\bprt Z^{(1)}\zbz}{\prt Z^{(1)}\zbz}
&=& \mu_\bz^z\zbz +
 2\mu^{(2)}_\bz\zbz \frac{\partial^2 Z^{(1)}\zbz}{\partial
Z^{(1)}\zbz} \nn\\
&-& \partial\mu^{(2)}_\bz\zbz - \frac{4}{3}
\mu^{(2)}_\bz\zbz \partial\ln w\zbz\\
\mu_\bz^z(2 \zbz) \equiv \frac{\bprt Z^{(2)}\zbz}{\prt
Z^{(2)}\zbz} &=& \mu^z_\bz\zbz +
 2\mu^{(2)}_\bz\zbz \frac{ \partial^2 Z^{(2)}\zbz}{\partial
Z^{(2)}\zbz } \nn\\
&-& \partial \mu^{(2)}_\bz\zbz - \frac{4}{3}
\mu^{(2)}_\bz\zbz  \partial\ln w\zbz, 
\lbl{beltramiw3}
\end{eqnarray}
and the Bilal-Fock-Kogan coefficients can be represented in terms
of the target Laguerre-Forsyth coordinates as:
\begin{eqnarray}
\mu_\bz^{(2)}\zbz& =&{\dps \frac{\biggl(\frac{\bprt Z^{(2)}\zbz}{\prt
Z^{(2)}\zbz}-\frac{\bprt Z^{(1)}\zbz}{\prt
Z^{(1)}\zbz}\biggr)}{2\biggl(\frac{\prt^2 Z^{(2)}\zbz}{\prt
Z^{(2)}\zbz}- \frac{\prt^2 Z^{(1)}\zbz}{\prt Z^{(1)}\zbz} \biggr)}}
\lbl{mu2coord}
\\
\mu_\bz^z\zbz &=&\frac{\bprt Z^{(1)}\zbz}{\prt Z^{(1)}\zbz} - 2
\mu_\bz^{(2)}\zbz \frac{\prt^2 Z^{(1)}\zbz}{\prt
Z^{(1)}\zbz}+\prt\mu_\bz^{(2)}\zbz \nn\\
&& +\ \frac{4}{3}\mu_\bz^{(2)}\zbz
\partial\ln w\zbz
\lbl{mu1coord}
\end{eqnarray}
(we point out to the reader's attention that on the real axis
$z=\bar{z}$ we have $\mu^{(2)}\zbz=0$ and $\mu^{(1)}\zbz=1$).
The  anomaly Eq\rf{wellanomaly} can be written in the $\cW(3)$
case as \cite{Garajeu:1995jn}:
 \begin{eqnarray}
&&\delta_\cW\Gamma_s= \int\Delta^{1}_{2}\zbz =\int\biggl\{\biggl(\cC\zbz
L_3\mu^z_\bz\zbz-\mu^z_\bz\zbz L^3\cC\zbz\biggr)\nn\\
& -&\frac{1}{3}\biggl(\cC^{(2)}\zbz L_5\mu^{(2)}_\bz\zbz
-\mu^{(2)}_\bz\zbz
L^5\cC^{(2)}\zbz \biggr)\nn\\
 &-& 8\biggl(\cC^{(1)}\zbz \mu^{(2)}\zbz
-\mu^z_\bz\cC^{(2)}\zbz\biggr)\prt\cW\zbz\nn\\ 
 &-&24\cW\zbz\biggl(\cC^{(1)}\zbz\prt\mu^{(2)}_\bz\zbz 
- \mu^{(2)}_\bz\zbz\prt\cC\zbz  \nn\\
 &+&\cC^{(2)}\zbz \prt\mu^z_\bz\zbz
-\mu^z_\bz\zbz\prt\cC^{(2)}\zbz\biggr)\biggr\}dz\wedge d\bz,
\lbl{anomalygood}
 \end{eqnarray}
 where the fifth order $L_5$ Bol operator  reads:
\begin{eqnarray}
L_5 &=& \prt^5 + 10 \cT\zbz \prt^3 +15(\prt\cT\zbz)\prt^2 +
\biggl[ 9(\prt\cT\zbz) + 16 \cT^2\zbz \biggr]\prt \nn\\
 &+& 2\biggl[(\prt^3\cT\zbz) +8\cT\zbz (\prt\cT\zbz)\biggr].
\end{eqnarray}
Anyhow in this case  the projective connection $\cT\zbz$ is written
in terms of the coordinates as:
\begin{eqnarray}
\cT\zbz=\frac{1}{2} \prt^2 \ln w\zbz - \frac{1}{3} (\prt \ln
w\zbz )^2 + \frac{v\zbz }{w\zbz},
\lbl{Tw3}
\end{eqnarray}
while  the spin 3 tensor, $\cW\zbz$,  takes the expression:
\begin{eqnarray}
\cW\zbz &=& \frac{1}{24} \biggl(\frac{1}{2} \prt^3 \ln w\zbz
 - \prt^2 \ln w \zbz
\prt \ln w\zbz  - \frac{2}{9} (\prt \ln w\zbz )^3\biggr)\nn\\
& +&
  \frac{1}{16} \biggl( \frac{\prt v\zbz }{w\zbz }
  - \frac{5v\zbz }{3w\zbz }\prt \ln w\zbz  \biggr).
\lbl{Ww3}
\end{eqnarray}
The partial counterterm compensation trick of Eq\rf{scounterterm}
 can be repeated:
\begin{eqnarray}
&&\delta_\cW \int dz\wedge d\bz \Biggl[
\Biggl(\mu_\bz^z\zbz-\prt\mu_\bz^{(2)}\zbz\Biggr) \Biggl( \cT\zbz
-\zer{\cT}\zbz\Biggr)\nn\\
&& +\ 8\mu_\bz^{(2)}\zbz
\Biggl(\cW\zbz-\zer{\cW}\zbz\Biggr)\Biggr]=\nn\\
&&\int \Delta^1_2\zbz dz\wedge d\bz +\int dz\wedge d\bz \Biggl[
\delta_\cW\Biggl(\mu_\bz^z\zbz-\prt\mu_\bz^{(2)}\zbz\Biggr) \Biggl(
\cT\zbz -\zer{\cT}\zbz\Biggr)\nn\\
&& +\ 8 \biggl( \delta_\cW\mu_\bz^{(2)}\zbz
\biggr)\Biggl(\cW\zbz-\zer{\cW}\zbz\Biggr)\Biggr]
\lbl{countertermw3}
\end{eqnarray}
(where $\zer{\cT}\zbz$ and $\zer{\cW}\zbz$ are the components of the
background matrix $\zer{\cA}_z\zbz$ twined to the currents $\cT\zbz$
and $\cW\zbz$ respectively).

So we define a new quantum Action $\Gamma_Q$ as in Eq\rf{anomaly} by
\begin{eqnarray}
\Gamma_Q &=& \Gamma_S -\hbar\int dz\wedge d\bz \Biggl[
\Biggl(\mu_\bz^z\zbz-\prt\mu_\bz^{(2)}\zbz\Biggr) \Biggl( \cT\zbz
-\zer{\cT}\zbz\Biggr)\nn\\
&& +\ 8\mu_\bz^{(2)}\zbz \Biggl(\cW\zbz-\zer{\cW}\zbz\Biggr)\Biggr]
\lbl{GammaQw3}
\end{eqnarray}
with anomaly:
\begin{eqnarray}
&&\delta_\cW \Gamma_Q=\int dz\wedge d\bz \Biggl[
\delta_\cW\Biggl(\mu_\bz^z\zbz-\prt\mu_\bz^{(2)}\zbz\Biggr) \Biggl(
\cT\zbz -\zer{\cT}\zbz\Biggr)\nn\\
&&+\ 8 \biggl( \delta_\cW\mu_\bz^{(2)}\zbz
\biggr)\Biggl(\cW\zbz-\zer{\cW}\zbz\Biggr)\Biggr].
\lbl{qapw3}
\end{eqnarray}

The Ward identities can be obtained if we differentiate
 with respect $\cC^{(1)}\zbz$, and $\cC^{(2)}\zbz$~;
 so if we put
$<{\cX}>=0$ we get respectively:
\begin{eqnarray}
&&\left.
\biggl( \prt Z^{(1)}\zbz \frac{\delta \Gamma}{\delta
Z^{(1)}\zbz} +
 \prt Z^{(2)}\zbz \frac{\delta \Gamma}{\delta Z^{(2)}\zbz}
 \biggr)\right|_{<\cX>=0}\nn\\
 &=& \frac{\delta}{\delta \cC^{(1)}\zbz}\Biggl\{
\int d\zp\wedge d\bzp \Biggl[
\delta_\cW\Biggl(\mu_{\bzp}^{\zp}\zbzp-\prt'\mu_{\bzp}^{(2)}\zbzp\Biggr)
\Biggl(
\cT\zbzp -\zer{\cT}\zbzp\Biggr)\nn\\
& +&8 \biggl( \delta_\cW\mu_{\bzp}^{(2)}\zbzp
\biggr)\Biggl(\cW\zbzp-\zer{\cW}\zbzp\Biggr)\Biggr]\Biggr\}
\lbl{wardc1}
 \end{eqnarray}
and
\begin{eqnarray}
&&
\Biggl[\biggl(3 \prt^2 Z^{(1)}\zbz +\prt Z^{(1)}\zbz\prt
-\frac{4}{3} \prt Z^{(1)}\zbz \prt \ln w\zbz\biggr)\frac{\delta
\Gamma}{\delta
Z^{(1)}\zbz} \nn\\
&+&\left.\biggl(3 \prt^2 Z^{(2)}\zbz +\prt Z^{(2)}\zbz\prt -\frac{4}{3}
\prt Z^{(2)}\zbz \prt \ln w\zbz\biggr)\frac{\delta \Gamma}{\delta
Z^{(2)}\zbz}\Biggr]\right|_{<\cX>=0} \nn\\
&=&\frac{\delta}{\delta \cC^{(2)}\zbz}\Biggl\{ \int d\zp\wedge
d\bzp \Biggl[
\delta_\cW\Biggl(\mu_{\bzp}^{\zp}\zbzp-\prt'\mu_{\bzp}^{(2)}\zbzp\Biggr)
\Biggl(
\cT\zbzp -\zer{\cT}\zbzp\Biggr)\nn\\
& +&8 \biggl( \delta_\cW\mu_{\bzp}^{(2)}\zbzp
\biggr)\Biggl(\cW\zbzp-\zer{\cW}\zbzp\Biggr)\Biggr]\Biggr\}.
\lbl{wardc2}
\end{eqnarray}

Substituting  the functional derivative $\frac{\delta \Gamma}{\delta Z^{(2)}\zbz}$ in Eq \rf{wardc2} from
Eq\rf{wardc1}, surpisingly gives rise to
after some algebra, an algebraic (instead of differential)
equation for  $ \frac{\delta \Gamma}{\delta Z^{(1)}\zbz} $, namely
\begin{eqnarray}
&&\left.
\frac{\delta \Gamma}{\delta Z^{(1)}\zbz}\right|_{<{\cX}>=0}
= \frac{\hbar}{3 \left(\prt^2 Z^{(1)}\zbz-\frac{\prt
Z^{(1)}\zbz}{\prt Z^{(2)}\zbz} \prt^2 Z^{(2)}\zbz \right) -\prt
Z^{(2)}\zbz \prt \left( \frac{\prt Z^{(1)}\zbz}{\prt
Z^{(2)}\zbz}\right) }\nn\\
 && \int d\zp\wedge d\bzp
\Biggl[
\Biggl\{ \frac{\delta \Biggl( \delta_\cW
\left(\mu_{\bzp}^{\zp}\zbzp-\prt'\mu_{\bzp}^{(2)}
\zbzp\right) \Biggr)} {\delta \cC^{(2)}\zbz} -
\biggl(3 \prt^2 Z^{(2)}\zbz +\prt Z^{(2)}\zbz\prt_z  \nn\\
&& -\ \frac{4}{3} \prt Z^{(2)}\zbz \prt \ln
w\zbz\biggr) \frac{1}{\prt Z^{(2)}\zbz}\,
\frac{\delta\Biggl(\delta_\cW
\left(\mu_{\bzp}^{\zp}\zbzp-\prt'\mu_{\bzp}^{(2)}\zbzp\right) \Biggr)}
{\delta\cC^{(1)}\zbz} \Biggr\} \nn\\
&& \Biggl( \cT\zbzp -\zer{\cT}\zbzp\Biggr) + 8\,
\frac{\delta\biggl( \delta_\cW\mu_{\bzp}^{(2)}\zbzp
\biggr)}{\delta \cC^{(2)}\zbz}
\Biggl(\cW\zbzp-\zer{\cW}\zbzp\Biggr) \Biggr]
\lbl{gamma1}
\end{eqnarray}
and from Eq \rf{wardc1} it is possible to find a complicated
expression for  $ \frac{\delta \Gamma}{\delta Z^{(2)}\zbz} $.

These tadpoles are zero for $\cT\zbz=\zer{\cT}\zbz$, and
$\cW\zbz=\zer{\cW}\zbz$, but give rise to non zero Green functions when
further functional derivatives are performed. One has
\begin{eqnarray}
&& \left.
\frac{\delta \Gamma}{\delta Z^{(1)}(z'',\bar{z}'')\delta
Z^{(1)}\zbz}\right|_{\left\{\begin{array}{c}
\cT= \zer{\cT} \\
 \cW=\zer{\cW}\\
<\cX>=0\end{array}\right.} = \nn\\
&& \frac{\hbar}{3 \left(\prt^2 Z^{(1)}\zbz-\frac{\prt
Z^{(1)}\zbz}{\prt Z^{(2)}\zbz} \prt^2 Z^{(2)}\zbz \right) -\prt
Z^{(2)}\zbz \prt\left( \frac{\prt Z^{(1)}\zbz}{\prt
Z^{(2)}\zbz}\right)} \times \nn\\
 && \int d\zp\wedge d\bzp
\left[ \Biggl\{
\frac{\delta \Biggl( \delta_\cW
\left(\mu_{\bzp}^{\zp}\zbzp-\prt'\mu_{\bzp}^{(2)}
\zbzp \right) \Biggr)} {\delta \cC^{(2)}\zbz} -
\biggl(3 \prt^2 Z^{(2)}\zbz +\prt Z^{(2)}\zbz\prt_z \right. \nn\\
&& -\ \frac{4}{3} \prt Z^{(2)}\zbz \prt \ln w\zbz\biggr)
\frac{1}{\prt
Z^{(2)}\zbz}\,\frac{\delta\Biggl( \delta_\cW
\left(\mu_{\bzp}^{\zp}\zbzp-\prt'\mu_{\bzp}^{(2)}\zbzp \right) \Biggr)}
{\delta \cC^{(1)}\zbz} \Biggr\}\times \nn\\
&& \left.\left. \frac{\delta\cT\zbzp}{\delta
Z^{(1)}(z'',\bar{z}'')} + 8 \frac{\delta\biggl(
\delta_\cW\mu_{\bzp}^{(2)}\zbzp \biggr)}{\delta \cC^{(2)}\zbz}\,
\frac{\delta \cW\zbzp}{\delta
Z^{(1)}(z'',\bar{z}'')} \right]
\right|_{\left\{\begin{array}{c}\cT=\zer{\cT}\\
\cW=\zer{\cW}\\
\end{array}\right.} .
\lbl{gamma2}
\end{eqnarray}
Now if we exchange the functional derivative order sequence
$\zbz\lra (z'',\bar{z}'')$ we can evidentiate, with easy derivable (but
rather complicated) expressions,  the noncommutative character
of the complex coordinates considered as quantum fields.

\end{document}